\documentclass{article}

\usepackage[preprint]{neurips_2026}

\usepackage[utf8]{inputenc} 
\usepackage[T1]{fontenc}    
\usepackage{booktabs}       
\usepackage{amsfonts}       
\usepackage{nicefrac}       
\usepackage[final]{microtype}      
\usepackage[table]{xcolor}         
\usepackage{capt-of}
\usepackage{graphicx}
\usepackage{booktabs}

\usepackage[accsupp]{axessibility}  
\usepackage[breaklinks=true]{hyperref}
\usepackage{url}
\usepackage{booktabs}
\usepackage{amssymb}
\usepackage{algorithm}
\usepackage{siunitx}
\usepackage{algorithmic}
\usepackage{tabularx}
\usepackage{subcaption}
\usepackage{wrapfig}
\usepackage{caption}
\usepackage{ragged2e}
 \usepackage{tikz}
 \usepackage{array}     
\usepackage{adjustbox}
\usetikzlibrary{arrows.meta}
\usetikzlibrary{arrows.meta,positioning,fit,calc,shapes.misc}

\usepackage{bm}
\usepackage{colortbl}
\usepackage{graphicx} 
\usepackage{booktabs}
\usepackage{makecell}
\usepackage{multirow}
\usepackage[nameinlink,capitalise,capitalize]{cleveref}

\usepackage{siunitx}
\title{Any Orthogonal Transform Will Do: Range Reduction Explains Rotation-Based Quantization for Super-Resolution}

\author{%
  Dorsa Zeinali \quad Hailing Wang \quad Yitian Zhang \quad Yun Fu \\
  Khoury College of Computer Science, Northeastern University, Boston, MA, USA \\
  \texttt{\{zeinali.d, wang.haili, zhang.yitian, y.fu\}@northeastern.edu} \\
}

\begin{document}

\maketitle

\begin{abstract}
  Rotation-based quantization --- multiplying weights and activations by a Hadamard matrix before quantizing --- is now standard, but why it works is explained informally, via incoherence, the central limit theorem, or kurtosis, and rarely tested directly. We ask which property of the post-rotation distribution drives the gain, using paired hypothesis tests on the weights and activations of an image super-resolution transformer. Three candidate properties separate cleanly. \textbf{Range reduction} replicates under every orthogonal transform we tried --- Sylvester Hadamard, QuIP\#-Paley, DCT-II, and random orthogonal matrices --- and transfers to a state-space backbone and to CNN weights. \textbf{Improved normality} also replicates. The \textbf{increased mass near zero}, by contrast, is significant only under the zero-padded Sylvester construction and disappears under unpadded transforms, so we attribute it to the padding, not the transform. The end task follows the same pattern: over a grid of three scale factors, three bitwidths and five benchmarks, rotation improves PSNR in all $45$ configurations against an otherwise identical un-rotated quantizer at the same nominal bitwidth (the rotated arm carries $0.267$ bits/parameter of Sylvester padding at 4 bits and stays ahead in every cell when charged for it), while substituting DCT-II or a random orthogonal matrix costs only $0.06$--$0.07$\,dB, less than the padding the Hadamard is charged for. The choice of kernel is therefore an implementation decision: the Hadamard's advantage is its $\pm 1$ entries and $O(n\log n)$ butterflies. We package the recipe as CompSRT, which improves on 2DQuant in $44$ of $45$ configurations at equal nominal bitwidth --- charged for its padding that edge holds at 3 and 2 bits but not at 4. CondiQuant remains ahead at every operating point by $0.03$--$1.40$\,dB, which we report.
\end{abstract}

\section{Introduction}
\label{sec:intro}

Image super-resolution (SR) focuses on recovering fine-grained image details from low-resolution inputs and is widely used across applications such as image and video enhancement \cite{hitachisuper,takeda2009super,su2011spatially}, medical imaging \cite{yu2017computed,greenspan2002mri,yu2018super,robinson2017new}, and remote sensing \cite{murthy2014skysat, zhu2018spatio}. Convolutional neural networks (CNNs) like Enhanced Deep Residual Networks (EDSR) \cite{lim2017enhanced}, Residual Dense Networks (RDN) \cite{zhang2018residualdensenetworkimage}, and Super-Resolution Residual Network (SRResNet) \cite{ledig2017photo} have achieved high performance in these tasks but have high parameter counts. Transformer-based models like SwinIR-light have emerged as efficient alternatives, offering competitive performance with fewer parameters. 

Despite being lighter than traditional CNNs, SwinIR-light still demands significant resources. 
Model compression addresses this issue, most directly through quantization. Quantization works by reducing parameter bitwidths (e.g., from 32/16 to 2/4-bit), either during quantization-aware training (QAT), which jointly optimizes weights and quantization parameters, or post-training quantization (PTQ), where quantization parameters are calibrated on a frozen model.

While most prior work focuses on CNNs, recent PTQ methods like 2DQuant~\citep{liu20242dquant} and CondiQuant~\citep{liu2025condiquant} have adapted PTQ to SwinIR. Both still leave a gap to the full-precision (FP) SwinIR-light model, and neither asks which property of the weight and activation distributions a quantizer should be shaping --- a question the LLM-quantization literature answers only informally.

Prior work on LLM quantization~\citep{ashkboos2024quarot,liu2024spinquant,federici2025hadanorm,chee2023quip,sun2024flatquant,tseng2024quip} has found that outliers in weights and activations cause performance degradations, and that Hadamard transformations help by increasing the ``flatness'' of distributions~\citep{sun2024flatquant,federici2025hadanorm,tseng2024quip}. The mechanism explanations differ across these works: \citet{tseng2024quip} prove incoherence bounds showing that Hadamard transforms concentrate entry magnitudes; \citet{liu2024spinquant} argue that random rotations blend large and small weights into a well-behaved distribution and analyze activations via kurtosis; \citet{federici2025hadanorm} appeal to the central limit theorem. These mechanism claims are made informally or via single summary statistics. To our knowledge, no prior work \emph{jointly} tests, on both weights and activations, whether the transform statistically significantly changes range, near-zero mass, and normality with paired hypothesis tests and effect sizes. We provide that test. Our contributions are as follows.
\begin{list}{\labelitemi}{%
  \setlength{\leftmargin}{1.1em}%
  \setlength{\labelwidth}{0.7em}%
  \setlength{\labelsep}{0.4em}%
  \setlength{\itemindent}{0pt}%
  \setlength{\listparindent}{0pt}%
  \setlength{\rightmargin}{0pt}%
  \setlength{\topsep}{3pt}%
  \setlength{\partopsep}{0pt}%
  \setlength{\parsep}{0pt}%
  \setlength{\itemsep}{3pt}%
}
\item \textbf{Which distributional property the gain comes from.} We give a paired statistical decomposition, on the weights and activations of an SR transformer, of why orthogonal transforms aid quantization. Range reduction is the primary driver and it is kernel-independent: $d_z{=}0.43$ on weights and $0.95$ on activations under Sylvester+padding, $0.36$--$0.42$ and $0.51$--$0.85$ under QuIP\#-Paley, DCT and random-orthogonal substitutes, and $0.99$ on EDSR conv weights, with $p<10^{-6}$ throughout. Normality improvement is kernel-independent as well. The $\varepsilon$-band increase is not: it is significant only under Sylvester+padding ($d_z{=}2.28$) and disappears under unpadded Paley, DCT and random-orthogonal kernels, so we attribute it to the zero-padding step, not the transform (\Cref{sec:padded-defense}).
\item \textbf{An additive scalar reparameterization, and what it is worth.} We reparameterize the quantization scalar and zero offset additively as $S{+}\alpha$ and $l{+}\beta$, decoupling the gradient pathways for scale and offset at finetune time without re-deriving LSQ-style step-size learning. Ablating the decomposition (\Cref{tab:scalar-ablation}) attributes the accuracy to bound learning and to the rotation, in that order. The additive correction itself contributes no measurable gain, and replacing it with direct step-size learning changes nothing (\Cref{tab:methodology-ablations}); we report this rather than claim the component works.
\item \textbf{The rotation's end-task contribution, charged for its own padding.} Over a uniform grid of three scale factors, three bitwidths and five benchmarks, the Hadamard improves PSNR in all $45$ configurations relative to an otherwise identical un-rotated quantizer, with the gain generally growing as the budget tightens: a mean of $+0.20$\,dB at 4-bit $\times 2$ rising to $+0.34$\,dB at 2-bit $\times 4$. The rotated arm carries the Sylvester zero-padding --- $0.267$/$0.200$/$0.133$ more bits per parameter at $4$/$3$/$2$ bits --- and charged for that at this model's own bit-rate the advantage narrows to $+0.02$--$0.20$\,dB, but stays positive in all nine (scale, bit) cells (\Cref{sec:bitcharge}). This is the end-task counterpart of the distributional analysis above, and it separates the rotation from the rest of an SR PTQ recipe.
\item \textbf{A like-for-like comparison, including the gap we do not close.} Against prior PTQ for SwinIR-light on the same grid, CompSRT improves on 2DQuant in $44$ of $45$ configurations (up to $+1.08$\,dB, with one tie), significant at $\alpha{=}0.05$ in $8$ of the $9$ (scale, bit) cells. Both figures are at equal nominal bitwidth; charged for the Sylvester padding the advantage is positive in $7$ of $9$ cells and significant in $5$, failing at 4 bits and holding at 3 and 2 (\Cref{sec:bitcharge}). CondiQuant remains ahead of CompSRT at every operating point, by $0.03$--$1.40$\,dB. We report that gap rather than omit it, and note that it widens monotonically as the bit budget tightens, from a mean of $0.12$\,dB at 4-bit to $0.24$\,dB at 3-bit and $0.65$\,dB at 2-bit.
\end{list}
\section{Related Work}
\label{sec:related}
\subsection{Image Super-Resolution}
EDSR (Enhanced Deep Super-Resolution) \cite{lim2017enhanced}, is a CNN-based architecture that improves upon traditional residual networks by removing unnecessary modules and stabilizing the training procedure. SRResNet \cite{ledig2017photo} employs residual learning and partial convolution based padding to generate high-quality images with fine details. SwinIR \cite{liang2021swinir} is a transformer-based model that has demonstrated superior performance with a reduced number of parameters compared to CNN-based approaches by leveraging shallow and deep feature extraction along with the self-attention mechanism. SwinIR is composed of Residual Swin Transformer Blocks (RSTBs), each of which contains multiple Swin Transformer Layers (STLs). SwinIR-light is the SwinIR model designed for lightweight SR made up of 4 RSTBs that each contain 6 STLs.
\begin{figure}[t]
\includegraphics[trim=55mm 80mm 48mm 85mm,clip,width=\linewidth]{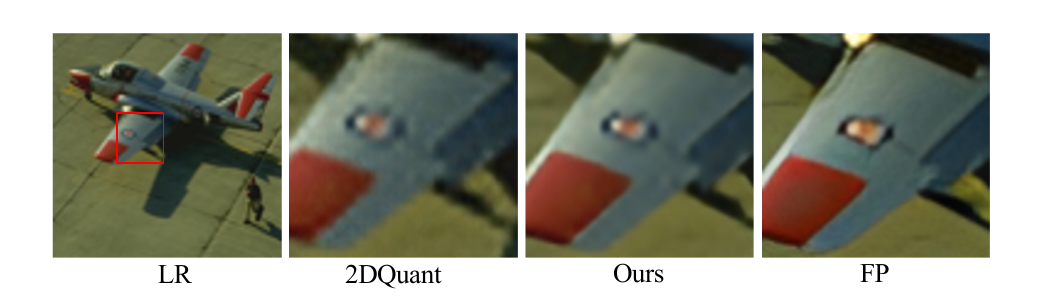}
\caption{Qualitative visual comparison for 2-bit $(\times4)$ SR on a challenging example. LR denotes low resolution image, FP denotes the output of the FP model. The comparative example is taken from 2DQuant~\citep{liu20242dquant}\protect\footnotemark[1]. The prior open-source baseline (2DQuant) is visibly blurrier than our output on this example.}
\vspace{-3mm}
\label{fig:figure_2}
\end{figure}
\footnotetext[1]{Since CondiQuant does not have open-sourced code, we compare our visual results with 2DQuant.}
\subsection{Model Quantization}

 Previous research has focused mainly on PTQ for CNN based architectures or vision transformers \cite{hong2024adabm,makhov2024towards,tu2023toward,ding2022towards,yuan2024ptq4vitposttrainingquantizationvision} \mbox{\cite{li2023repqvitscalereparameterizationposttraining,liu2023noisyquantnoisybiasenhancedposttraining}}, QAT \cite{tian2023cabm,hong2024overcoming,hong2022cadyq,hong2022daq,li2020pams,zhong2022dynamic,wang2021fullyquantizedimagesuperresolution} or unique quantized architectures \cite{qin2023quantsr}. 2DQuant~\citep{liu20242dquant} and CondiQuant~\citep{liu2025condiquant} represent the SOTA PTQ methods for SwinIR-light. 2DQuant performs Distribution-Oriented Bound Initialization (DOBI) to search for optimal clipping ranges from input distributions, then finetunes these bounds on a calibration dataset to minimize discrepancy with the full precision model's output. CondiQuant identifies that quantization errors primarily stem from activation quantization and uses the condition number of weight matrices to measure how sensitive outputs are to small input changes, employing proximal gradient descent to minimize these condition numbers while preserving model outputs. 
While both methods are effective, they do not directly address large ranges in weight and activation distributions. 2DQuant initializes parameters based on distributions but doesn't modify them, while CondiQuant focuses on condition numbers rather than distribution properties. Our method goes further by shrinking the ranges of the weights and activations and compacting the information signals through Hadamard transforms, making distributions more quantization-friendly.

\noindent\textbf{Hadamard- and rotation-based quantization in LLMs.} QuaRot~\citep{ashkboos2024quarot} and SpinQuant~\citep{liu2024spinquant} apply learned or fixed rotations to suppress activation outliers in large language models; HadaNorm~\citep{federici2025hadanorm} cites the central limit theorem; QuIP and QuIP\#~\citep{chee2023quip,tseng2024quip} motivate the transform via incoherence theorems; FlatQuant~\citep{sun2024flatquant} frames the goal as ``flatness'' measured by kurtosis. These works share the technique but differ in two respects from our setting. First, none disaggregate \emph{which} property of the post-Hadamard distribution drives the quantization gain, nor jointly verify the effect on both weights and activations with paired hypothesis tests. Second, their analyses target billion-parameter LLMs where compute is dominated by memory movement and FP32-rotation overhead is negligible; in the small SR-transformer regime, parameter and runtime budgets do not absorb the same overhead, which we quantify in \Cref{sec:complexity}. We treat the empirical decomposition as the contribution prior work has left unanswered. The Hadamard is preferred over other orthonormal transforms (DCT/DST/random orthogonal) because all preserve L2 energy --- so our range and $\varepsilon$-band arguments transfer --- but only Hadamard admits a multiplication-free $O(n\log n)$-addition implementation with $\pm 1$ entries, avoiding stored coefficients; prior LLM-quantization work selects Hadamard for the same reasons~\citep{ashkboos2024quarot,tseng2024quip}.

\noindent\textbf{LSQ / LSQ+ contrast.} LSQ~\citep{esser2019learned} and LSQ+~\citep{bhalgat2020lsq} learn the quantization step size (and offset) directly. We do not learn $S$ or $l$ directly; we learn the bounds $u,l$ (as in 2DQuant) and add small additive corrections $\alpha,\beta$ that decouple $S$ and $l$ at finetune time. The distinction matters for the bit budget as well as for optimisation. Because the integer code is clamped to $[0,2^b{-}1]$ (\Cref{eq:codeclamp}), the quantizer emits exactly $2^b$ levels whatever $\alpha$ does; $\alpha$ rescales the step, and so sets how widely those levels are spread rather than how many there are. In the trained models this narrowing is substantial: at 2-bit the median representable span is $0.71$ of the calibrated range for weight quantizers and $0.82$ for activation quantizers, with $\alpha<0$ in essentially every quantizer. The correction therefore acts as a learned tightening of the calibrated clipping range, and it settles at nearly the same place with and without the rotation (activations: $0.82$ versus $0.82$; weights: $0.71$ versus $0.69$). \Cref{tab:scalar-ablation} isolates its contribution against the same recipe with the correction held inert.

\noindent\textbf{Quantization-aware training (QAT) baselines.} QAT methods such as QuantSR~\citep{qin2023quantsr} and ODM jointly retrain weights and quantization parameters end-to-end, typically attaining higher accuracy at the cost of orders-of-magnitude more compute and access to full training data. PTQ targets the deployment regime where this is unavailable. We compare against PTQ baselines because they share our deployment constraints; we treat QAT methods as a complementary upper bound rather than a fair head-to-head comparison.

\section{Methodology}
\subsection{Hadamard Transformation}\label{sec:hadamard}
\begin{figure*}[t]
  \centering
\includegraphics[trim=3.9mm 4mm 0mm 4mm,clip, width=\linewidth]{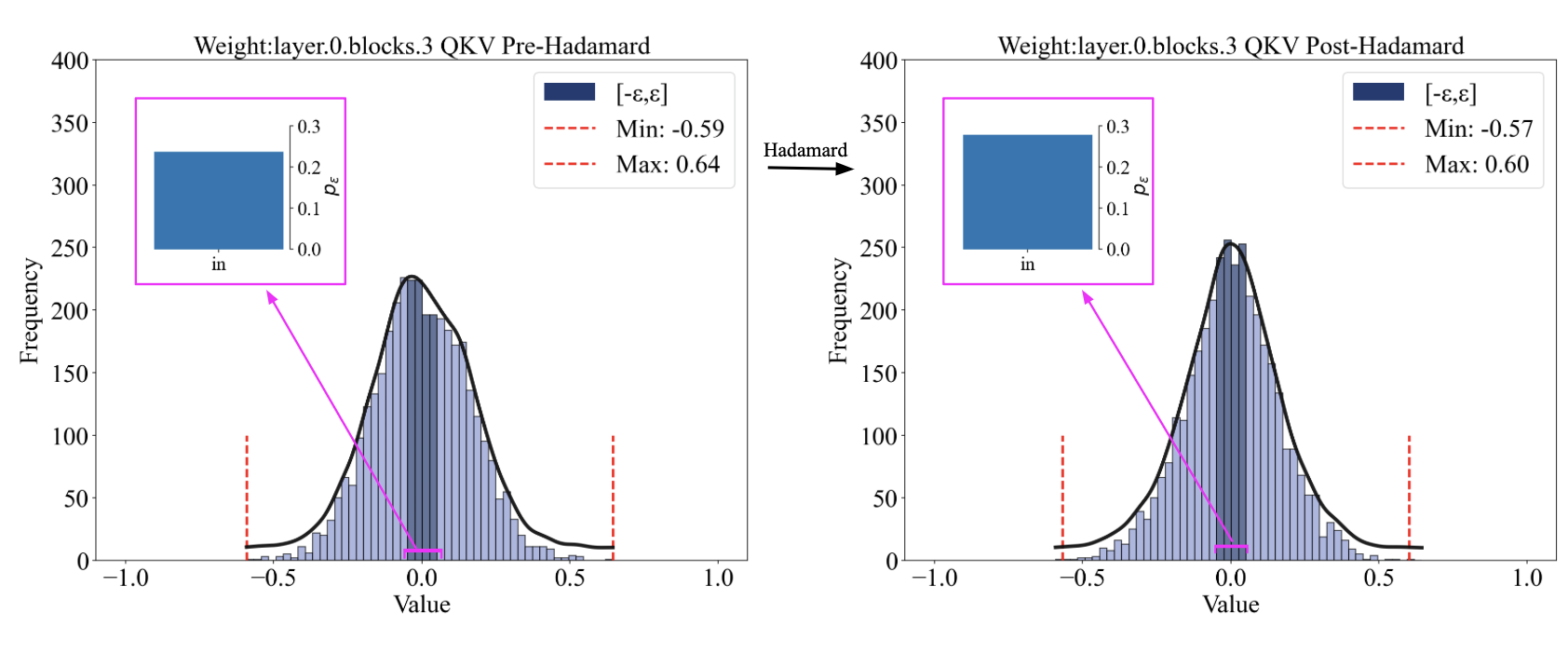}
  \caption{The left histogram shows the weight distribution prior to the Hadamard with a larger range and a sharper peak. The right histogram shows the post-Hadamard distribution is more Gaussian with smaller dynamic range. The dark blue band indicates values within $[-\varepsilon,\varepsilon]$ at $\varepsilon=0.05$ and $p_\varepsilon$ is the in-band fraction. This visualization uses the Sylvester+padding kernel; the range reduction shown here replicates under both the Sylvester and the QuIP\#-Paley constructions, while the magnitude of the $p_\varepsilon$ increase is partially construction-dependent (see \Cref{tab:transform-ablation}).}
\label{fig:hadamard_transformation}
\vspace{-5mm}
\end{figure*}

Hadamard transformations have been used effectively in LLM quantization to reduce outliers and make matrices tend towards the Gaussian, but how the transform actually functions has remained underexplored. Hadamard matrices have entries in $\{\pm 1\}$ and define linear orthogonal transforms; the Sylvester construction~\citep{sylvester1867thoughts} yields them for power-of-two dimensions, and the Paley~\citep{paley1933orthogonal} and Williamson~\citep{williamson1944hadamard} constructions extend the family to a wide set of multiples of four that are not powers of two (e.g., orders $12, 20, 28, 36, 44, 52, 60, \dots$). The QuIP\#-Paley kernel of \citet{tseng2024quip} composes the seed Hadamards available in its released table with Sylvester butterflies to handle any dim of the form $K\cdot 2^k$ for $K$ in that table. For SwinIR-light's embedding dims ($60, 120, 180$), this avoids zero-padding entirely for $60$ and $120$ and pads only $180\to 192=12\cdot 16$ in the current QuIP\# seed table (a $180\times 180$ Hadamard via direct Paley-I exists, since $179$ is prime with $179\equiv 3\pmod 4$, but is not in the released codebase), versus a Sylvester-only implementation that pads to $64,128,256$. We apply $X'=(XH^\top)/\sqrt{n}$ along the last (channel) dimension in full precision; since $H$ is orthogonal, $\|XH^\top\|_2^2=\|X\|_2^2$, so the range statistic is robust to the construction while the $\varepsilon$-band statistic depends on it (\Cref{sec:padded-defense}).

\begin{figure*}[!t]
  \centering
  \begin{minipage}[t]{0.49\textwidth}
    \vspace{0pt} 
    \centering
    \includegraphics[width=\linewidth]{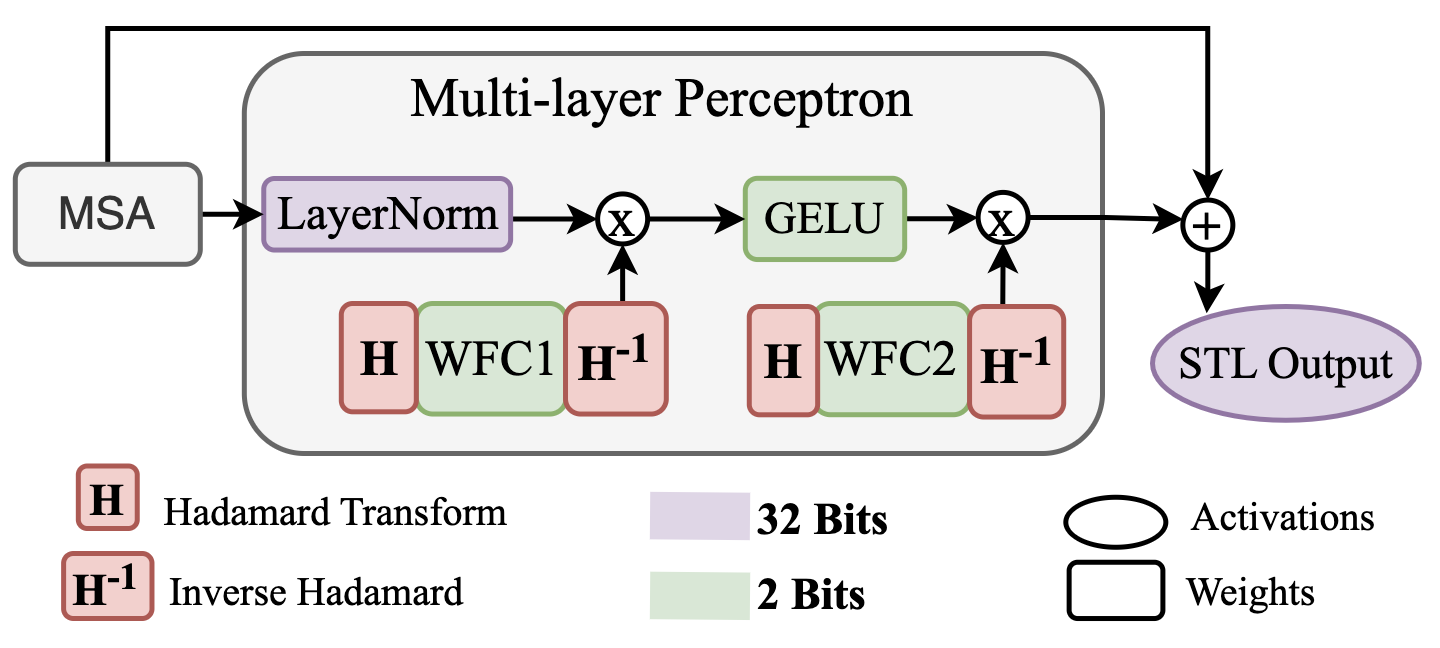}
    \caption{Architecture and quantization scheme for STL MLP; the Hadamard transformation runs in full precision while the surrounding linear layers operate at low bit-width.}
    \label{fig:stlmlp}
  \end{minipage}\hfill
  \begin{minipage}[t]{0.49\textwidth}
    \newcolumntype{Y}{>{\centering\arraybackslash}X}
\newcommand{\pval}[2]{\ensuremath{#1 e^{#2}}}

    
\vspace{2pt}
\centering
\small
\setlength{\tabcolsep}{1.5pt}
\renewcommand{\arraystretch}{1.3}

\begin{adjustbox}{max width=\linewidth}
\begin{tabular}{@{}|c|
>{\centering\arraybackslash}c@{\hspace{4pt}}>{\centering\arraybackslash}c|
>{\centering\arraybackslash}c@{\hspace{4pt}}>{\centering\arraybackslash}c|@{}}
\hline
\multirow[c]{2}{*}{\centering Test}
& \multicolumn{2}{c|}{Weights}
& \multicolumn{2}{c|}{Acts} \\
\cline{2-5}
& $p$ & $d_z$ & $p$ & $d_z$ \\
\hline
$\tilde{\Delta_w} > 0$
& \pval{2.3}{-8} & $0.39$
& \pval{3.3}{-39} & $0.36$ \\
\hline
$\tilde{\Delta_r} > 0$
& \pval{8.7}{-13} & $0.43$
& \pval{2.0}{-35} & $0.95$ \\
\hline
$\tilde{\Delta_p} > 0$
& \pval{4.0}{-25} & $2.28$
& \pval{1.0}{-4} & $0.61$ \\
\hline
\end{tabular}
\end{adjustbox}
    \captionof{table}{Combined statistical tests on the SwinIR-light $\times 2$ pre/post-Hadamard tensors. $\tilde{\Delta}_w, \tilde{\Delta}_r, \tilde{\Delta}_p$ are median paired deltas for normality (Shapiro--Wilk $W$), range, and $\varepsilon$-band proportion. One-sided Wilcoxon $p$-values; Cohen's $d_z$. $N_{\text{w}}{=}144$, $N_{\text{a}}{=}240$, all $p<0.05$.}
    \label{tab:wtest-combined}
  \end{minipage}
\end{figure*}

We test three properties on the SwinIR-light $\times 2$ model. Because tensor elements are not individually matchable across the Hadamard (each post-entry mixes many pre-entries), we treat each whole tensor as the experimental unit and form paired pre/post summaries: $N=144$ weight pairs and $N=240$ activation pairs (\Cref{sec:stats-repro} reproduces all six effect sizes from a fresh calibration pass). For each property we use a one-sided Wilcoxon signed-rank test with ties dropped on the paired deltas $\{\Delta_j\}$, $H_0{:}\,\mathrm{median}(\Delta){=}0$ vs.\ $H_1{:}\,\mathrm{median}(\Delta){>}0$, and report Cohen's $d_z$ (\Cref{tab:wtest-combined}; $\tilde{\Delta}$ denotes the median). For \emph{normality}, we randomly sample $10^6$ elements per tensor, compute the Shapiro--Wilk $W$ statistic, and form $\Delta_{wj}=W_{qj}-W_{pj}$. Distributions become statistically significantly more normal after the transformation, also visible in \Cref{fig:hadamard_transformation}.

For \emph{range}, we right-pad the pre-tensor with zeros to the post length so summaries live in the same ambient space, then form $\Delta_{rj}=R^{\text{pre}}_j-R^{\text{post}}_j$ where $R = \max-\min$. Both weights and activations show statistically significant range reduction at $\alpha=0.05$. This effect is not Hadamard-specific: DCT-II and random orthogonal substitutes give similar effect sizes ($d_z\in[0.36,0.85]$, $p<10^{-6}$; \Cref{tab:transform-ablation}), and it transfers to EDSR conv weights ($d_z{=}0.99$, $N{=}35$; \Cref{sec:edsr-stats}).

For the \emph{$\varepsilon$-band proportion}, we form $\Delta_{pj}=\hat p^{\text{post}}_j-\hat p^{\text{pre}}_j$ where $\hat p_j=\tfrac{1}{n_j}\sum_i \mathbf{1}\{|x_{j,i}|\le\varepsilon\}$ at $\varepsilon=0.05$ (sensitivity in \Cref{sec:epssensitive}). The Sylvester+padding implementation reported in \Cref{tab:wtest-combined} shows a significant increase ($d_z{=}2.28/0.61$ for weights/acts), but the magnitude of this effect depends on the Hadamard construction: under the QuIP\#-Paley kernel that does not zero-pad dim $60$ or $120$, the increase is reduced and not significant (\Cref{sec:padded-defense}, \Cref{tab:transform-ablation}). Range reduction, in contrast, replicates under both constructions with similar effect sizes ($0.43/0.95$ and $0.38/0.85$ for weights/acts), so we treat range reduction as the primary kernel-independent mechanism, with normality improvement as a kernel-independent secondary effect; the in-band-mass effect is significant under Sylvester+padding but partially construction-dependent.
\begin{figure*}[t]
\centering
\includegraphics[trim=0mm 1mm 4mm 1mm,clip,width=0.78\textwidth]{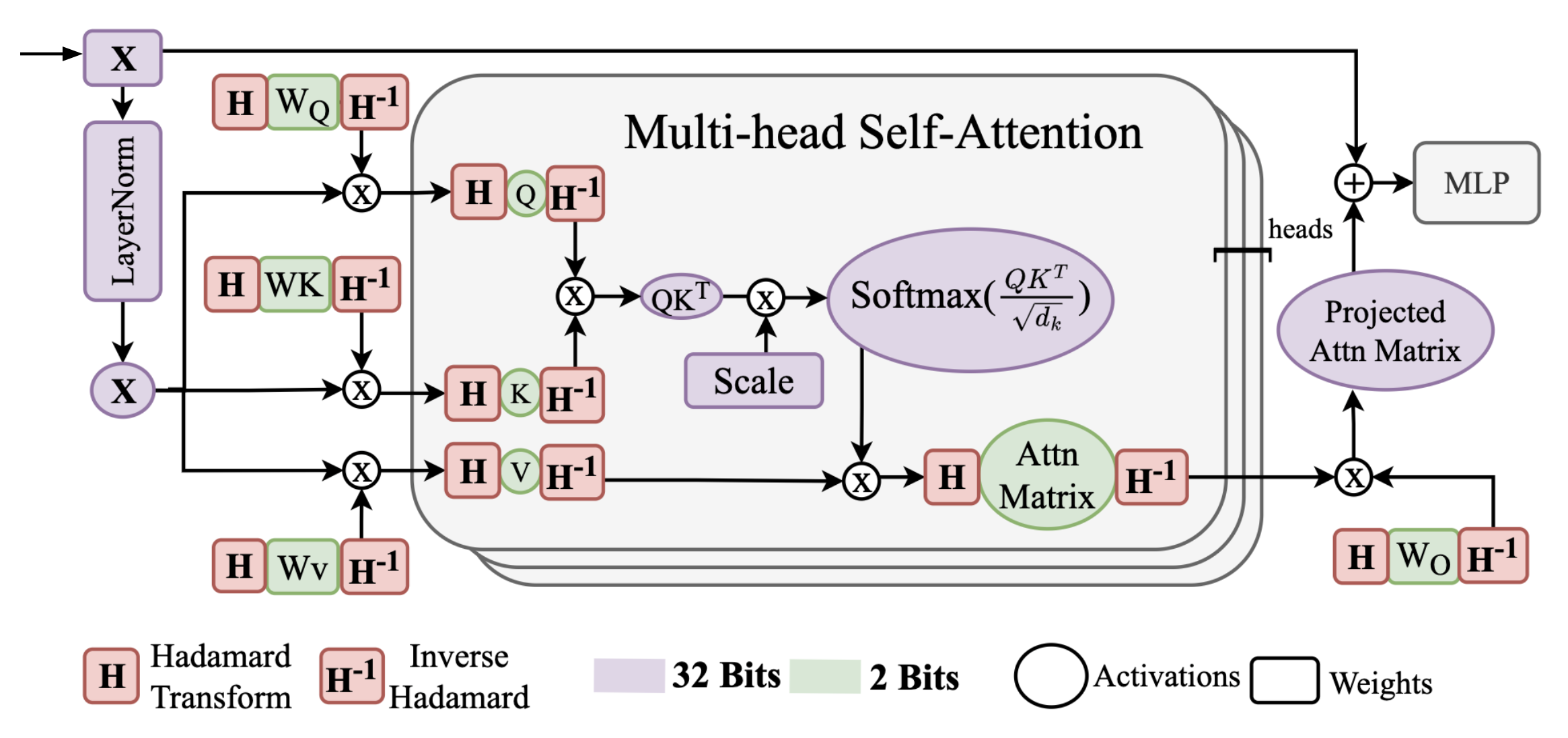}
\caption{Architecture and  Quantization scheme for Swin Transformer Layer (STL). X denotes the input. The quantized weights \& activations per component are in green and the FP operations are in purple. The Hadamard and inverse Hadamard transforms are shown with red boxes.}
\label{fig:stl}
\end{figure*}

\noindent\textbf{Why range reduction and not normality.} Only the $\varepsilon$-band candidate is eliminated: range reduction and normality both replicate under every kernel, so the claim needs a reason to prefer one. Two facts point the same way. The properties differ sharply on activations, where the rotation reduces range with $d_z{=}0.95$ but improves normality with only $d_z{=}0.36$, against $0.43$ and $0.39$ on weights (\Cref{tab:wtest-combined}); and the end-task benefit is entirely activation-side, since rotating activations alone is worth $+0.36$\,dB while rotating weights alone \emph{costs} $0.22$\,dB (\Cref{sec:rotation-scope}). Range reduction separates the two operands in the direction the end-task result requires; normality does not. We read it as the driver without claiming the two are cleanly separable: a rotation that compresses a heavy-tailed range also makes the tensor look more Gaussian, and these tests cannot rule out that normality is carried along.

\subsection{Scalar Decomposition}
We quantize attention layers, linear layers and batch matrix multiplication weights and activations. The full pipeline of our method and what is quantized is shown in \Cref{fig:stl} and \Cref{fig:stlmlp}. To quantize the parameters to 2-4 bits, we use fake-quantization \cite{jacob2017quantizationtrainingneuralnetworks} i.e. quantization dequantization that simulates the information degradation of the quantization process by restricting the values to be representable by 2-4 bits but then immediately reverting them back to floats. The standard approach to fake quantization involves determining the quantization scalar, $ S = \frac{\mathrm{u} - \mathrm{l}}{2^b - 1}$, based on the upper and lower bounds of the clipping range. The equations are as follows: 
\begin{equation}
S=\tfrac{u-l}{2^b-1},\quad
v_c=\mathrm{clip}(x,l,u),\quad
v_q=\mathrm{round}\!\left(\tfrac{v_c-l}{S}\right),\quad
v_{\text{deq}}=S\,v_q+l,
\label{eq:fakeq}
\end{equation}
where $x$ is the tensor to be quantized, $b$ is the bit-width, $l$ the lower bound (zero offset), and $u$ the upper bound. The lower bound makes zero exactly representable in the quantized range. Following 2DQuant, we first search for $u,l$ to minimize quantization error and then finetune them on calibration data. We additively decompose the scalar and zero offset as $S' = S+\alpha$ and $l'=l+\beta$, introducing two additional learnable parameters per layer.

The motivation is to decouple the gradient pathways for $S$ and $l$ at finetune time: in standard PTQ, $u$ and $l$ couple multiplicatively through $S=(u-l)/(2^b-1)$, so a step in either bound simultaneously shifts scale and offset. Adding $\alpha,\beta$ as small additive corrections introduces an independent search direction near the calibrated minimum without re-deriving LSQ-style step-size learning. Both are initialized at $0$ and learned during the finetuning phase together with $u,l$.

\noindent\textbf{Keeping the bit budget exact.} Because $\alpha$ rescales the step, the learned $S'$ is no longer tied to $(u-l)/(2^b-1)$, and an unconstrained integer code would let a nominally $b$-bit quantizer emit more than $2^b$ distinct values. We therefore clamp the code to the representable range,
\begin{equation}
v_q=\mathrm{clip}\!\left(\mathrm{round}\!\left(\tfrac{v_c-l'}{S'}\right),\,0,\,2^b-1\right),
\label{eq:codeclamp}
\end{equation}
so the quantizer emits exactly $2^b$ levels for every value of $\alpha$, and $b$ is a hard upper bound on the realised bitwidth rather than a nominal label. A negative $\alpha$ then narrows the representable span to $(2^b-1)S' < (u-l)$, which makes the correction a form of learned clipping in the sense of PACT~\citep{choi2018pact} and LSQ~\citep{esser2019learned}: the quantizer spends its levels over a tighter range and saturates the tail. We verify the resulting bitwidth directly in \Cref{sec:bitaudit}. The ablation in \Cref{sec:scalarablation} isolates what $\alpha,\beta$ contribute under this constraint.
\definecolor{lav32}{HTML}{E4DDF7}   
\definecolor{mint2}{HTML}{CFEAD0}   
\definecolor{brick}{HTML}{E8A69A}   
\definecolor{ink}{HTML}{3A3A3A}     
\tikzset{
  stage/.style={draw=ink, rounded corners, align=center, inner sep=4pt,
                minimum height=9mm, minimum width=28mm, font=\footnotesize},
  stage32/.style={stage, fill=lav32},
  stageQ/.style={stage, fill=mint2},
  flowarrow/.style={-{Latex[length=2.7mm]}, thick, draw=ink},
  hbadge/.style={draw=ink, fill=brick, rounded corners=2pt,
                 inner sep=1.2pt, font=\scriptsize\bfseries},
}

\section{Experiments}
\subsection{Data and Evaluation} 
We use DF2K \cite{8014883,lim2017enhanced} as the training data, which combines DIV2K \cite{8014883} and Flickr2K \cite{lim2017enhanced}. We use the Set5 \cite{bevilacqua2012low} as the validation set. We test our method on five commonly used benchmarks in the SR field: Set5 \cite{bevilacqua2012low}, Set14 \cite{10.1007/978-3-642-27413-8_47}, B100 \cite{937655}, Urban100 \mbox{\cite{7299156}}, and Manga109 \cite{matsui2017sketch}. The evaluation metrics we used are peak signal-to-noise ratio (PSNR) and structural similarity index measure (SSIM) \cite{1284395}. For both metrics, higher indicates better performance. The implementation details of our method are given in \Cref{sec:impdetails}. 
\begin{table*}[t]
\centering
\small
\setlength{\tabcolsep}{3pt}
\renewcommand{\arraystretch}{1}
\begin{tabularx}{\textwidth}{l|c|XX|XX|XX|XX|XX}
\hline

\multirowcell{2}[-0.5ex][c]{{Method ($\times$4)}} &
\multirowcell{2}[-0.5ex][c]{{Bit}} &

\multicolumn{2}{c|}{{Set5 }} & 
\multicolumn{2}{c|}{{Set14 }} & 
\multicolumn{2}{c|}{{B100}} & 
\multicolumn{2}{c|}{{Urban100}} & 
\multicolumn{2}{c}{{Manga109}} 
\\
& & {PSNR} & {SSIM} & {PSNR} & {SSIM} & {PSNR} & {SSIM} & {PSNR} & {SSIM} & {PSNR} & {SSIM} \\
\hline
SwinIR-light & 32 & 32.45 & 0.898 & 28.77 & 0.786 & 27.69 & 0.741 & 26.48 & 0.798 & 30.92 & 0.915 \\
Bicubic & 32 & 27.56 & 0.790 & 25.51 & 0.682 & 25.54 & 0.647 & 22.68 & 0.635 & 24.19 & 0.767 \\
\noalign{\vskip 1pt}\hline\noalign{\vskip 1pt}
PTQ4ViT & 4 & 31.49 & 0.883 & 28.04 & 0.768 & 27.20 & 0.724 & 25.53 & 0.766 & 29.52 & 0.894 \\
NoisyQuant & 4 & 31.09 & 0.875 & 27.75 & 0.759 & 26.91 & 0.715 & 25.07 & 0.750 & 28.96 & 0.882 \\
2DQuant & 4 & 31.77 & 0.887 & 28.30 & 0.773 & 27.37 & 0.728 & 25.71 & 0.771 & 29.71 & 0.897 \\
CondiQuant & 4 &  \textbf{32.09} &  \textbf{0.892} &  \textbf{28.50} &  \textbf{0.779} &  \textbf{27.52} &  \textbf{0.735} &  \textbf{25.97} &  \textbf{0.783} &  \textbf{30.16} &  \textbf{0.905} \\
CompSRT w/o Had. & 4 & 31.74 & 0.886 & 28.26 & 0.772 & 27.34 & 0.727 & 25.61 & 0.767 & 29.60 & 0.896 \\
CompSRT & 4 & 31.98 & 0.890 & 28.45 & 0.777 & 27.47 & 0.732 & 25.83 & 0.776 & 29.95 & 0.901 \\
\noalign{\vskip 1pt}\hline\noalign{\vskip 1pt}
PTQ4ViT & 3 & 29.77 & 0.834 & 27.00 & 0.725 & 26.21 & 0.674 & 24.22 & 0.698 & 27.94 & 0.848 \\
NoisyQuant & 3 & 28.90 & 0.797 & 26.50 & 0.697 & 26.16 & 0.663 & 23.86 & 0.667 & 27.17 & 0.812 \\
2DQuant & 3 &  {30.90} &  {0.870} &  {27.75} &  {0.757} &  {26.99} &  {0.713} &  {24.85} &  {0.736} &  {28.21} &  {0.868} \\
CondiQuant & 3 &  \textbf{31.62} &  \textbf{0.886} &  \textbf{28.20} &  \textbf{0.772} &  \textbf{27.31} &  \textbf{0.727} &  \textbf{25.39} &  \textbf{0.762} &  \textbf{29.29} &  \textbf{0.892} \\
CompSRT w/o Had. & 3 & 31.07 & 0.874 & 27.84 & 0.760 & 27.02 & 0.715 & 24.88 & 0.739 & 28.33 & 0.871 \\
CompSRT & 3 & 31.40 & 0.880 & 28.04 & 0.766 & 27.18 & 0.721 & 25.15 & 0.750 & 28.91 & 0.882 \\
\noalign{\vskip 1pt}\hline\noalign{\vskip 1pt}
PTQ4ViT & 2 & 27.23 & 0.670 & 25.38 & 0.591 & 25.15 & 0.562 & 22.94 & 0.559 & 24.66 & 0.613 \\
NoisyQuant & 2 & 25.94 & 0.586 & 24.33 & 0.507 & 24.16 & 0.472 & 22.32 & 0.484 & 23.82 & 0.540 \\
2DQuant & 2 &  {29.53} &  {0.837} &  {26.86} &  {0.732} &  {26.46} &  {0.693} &  {23.84} &  {0.691} &  {26.07} &  {0.816} \\
CondiQuant & 2 & \textbf{30.64} & \textbf{0.867} & \textbf{27.59} & \textbf{0.757} & \textbf{26.93} & \textbf{0.714} & \textbf{24.54} & \textbf{0.728} & \textbf{27.67} & \textbf{0.861} \\
CompSRT w/o Had. & 2 & 29.68 & 0.840 & 26.93 & 0.734 & 26.50 & 0.695 & 23.90 & 0.694 & 26.16 & 0.819 \\
CompSRT  & 2  & 30.06 & 0.847 & 27.20 & 0.739 & 26.67 & 0.699 & 24.17 & 0.703 & 26.78 & 0.828 \\
\hline
\end{tabularx}
\caption{Performance comparison for scale factor $(\times4)$ across bit widths; best in each column in \textbf{bold}. Baseline results are taken from SwinIR-light~\citep{liang2021swinir}, PTQ4VIT~\citep{yuan2024ptq4vitposttrainingquantizationvision}, NoisyQuant~\citep{liu2023noisyquantnoisybiasenhancedposttraining}, 2DQuant~\citep{liu20242dquant}, and CondiQuant~\citep{liu2025condiquant}, as reported in their papers; CompSRT rows are measured here. ``CompSRT w/o Had.'' is the identical recipe with the Hadamard replaced by the identity at the same nominal bitwidth, and isolates the rotation's contribution; the rotated arm additionally carries the Sylvester zero-padding, so it stores $0.267$/$0.200$/$0.133$ more bits per parameter at $4$/$3$/$2$ bits. CompSRT improves on 2DQuant and on the un-rotated variant in every cell at this scale; CondiQuant leads both throughout, by $0.05$--$0.89$\,dB.}
\vspace{-3mm}
\label{tab:performance_comparison3}
\end{table*}
\subsection{Results} 
\Cref{tab:performance_comparison3} reports scale factor $(\times 4)$ at 2--4 bits; the $\times 2$ and $\times 3$ tables are in \Cref{sec:additional}. Together they form a uniform grid of $3$ scales $\times$ $3$ bitwidths $\times$ $5$ benchmarks, all trained and evaluated under the identical protocol of \Cref{sec:impdetails}.

\noindent\textbf{The rotation helps everywhere.} The clearest result is the comparison against ``CompSRT w/o Had.'', the same recipe with the Hadamard replaced by the identity at the same nominal bitwidth. CompSRT improves PSNR in \emph{all 45} configurations of the grid, with a minimum gain of $+0.08$\,dB and a mean that rises from $+0.20$\,dB at 4-bit $\times 2$ to $+0.34$\,dB at 2-bit $\times 4$. The rotated arm does carry the Sylvester zero-padding, $0.267$/$0.200$/$0.133$ more bits per parameter at $4$/$3$/$2$ bits; charged for that padding at this model's own bit-rate, the rotation's advantage narrows but remains positive in all nine (scale, bit) cells. Because the two arms differ only in the transform, this isolates the rotation's end-task value from the rest of the recipe, and it is consistent with the distributional analysis of \Cref{sec:hadamard}: the tighter the bit budget, the more range reduction is worth.

\noindent\textbf{Against prior PTQ.} CompSRT improves on 2DQuant in $44$ of the $45$ configurations, by up to $+1.08$\,dB (Manga109, 2-bit $\times 2$); the single exception is Urban100 at 4-bit $\times 2$, where the two tie at $31.84$\,dB. That comparison is at equal nominal bitwidth. Charged for the Sylvester padding, CompSRT's mean advantage over 2DQuant holds at 3 and 2 bits ($+0.04$ to $+0.37$\,dB) but goes negative at 4 bits at $\times 2$ and $\times 3$ (\Cref{sec:bitcharge}): the rotation buys accuracy where bits are scarce, not where they are cheap. CondiQuant remains ahead of CompSRT at every one of the $45$ operating points. The margin is small at 4 bits (mean $0.12$\,dB, as little as $0.03$\,dB on Set14 at $\times 2$) and grows as the budget tightens, to a mean of $0.24$\,dB at 3 bits and $0.65$\,dB at 2 bits, peaking at $1.40$\,dB on Manga109 at 2-bit $\times 3$. We report this rather than restrict the comparison grid: distribution shaping does not close the distance to conditioning-based PTQ at low bitwidths.

\noindent\textbf{Significance.} Each (scale, bit, metric) cell is tested with a one-sided paired Wilcoxon signed-rank test over the five benchmarks; the procedure is in \Cref{sec:results-significance}. As shown in \Cref{tab:binom_combined}, $8$ of the $9$ (scale, bit) cells reach $p=0.031$, the smallest value attainable at $N{=}5$; the exception is 4-bit $\times 2$, where the Urban100 tie leaves $N{=}4$ and $0.0625$ is the smallest $p$ available. These tests are on raw differences at equal nominal bitwidth; repeated on differences charged for the Sylvester padding, five of the nine cells remain significant (\Cref{sec:bitcharge}). With five benchmarks these tests establish consistency of sign, not the size of the effect.
Relative to full precision, the $\times 4$ 4-bit model gives up $0.47$\,dB on Set5 ($31.98$ vs.\ $32.45$) and the $\times 2$ 4-bit model $0.26$\,dB ($37.89$ vs.\ $38.15$); at 2 bits the deficits are $2.39$\,dB and $1.68$\,dB respectively. Closing the low-bit gap to full precision therefore remains open, and it is where the difference between PTQ methods is largest.

\subsection{Qualitative Results}
\begin{figure*}[t]
\includegraphics[width=\linewidth]{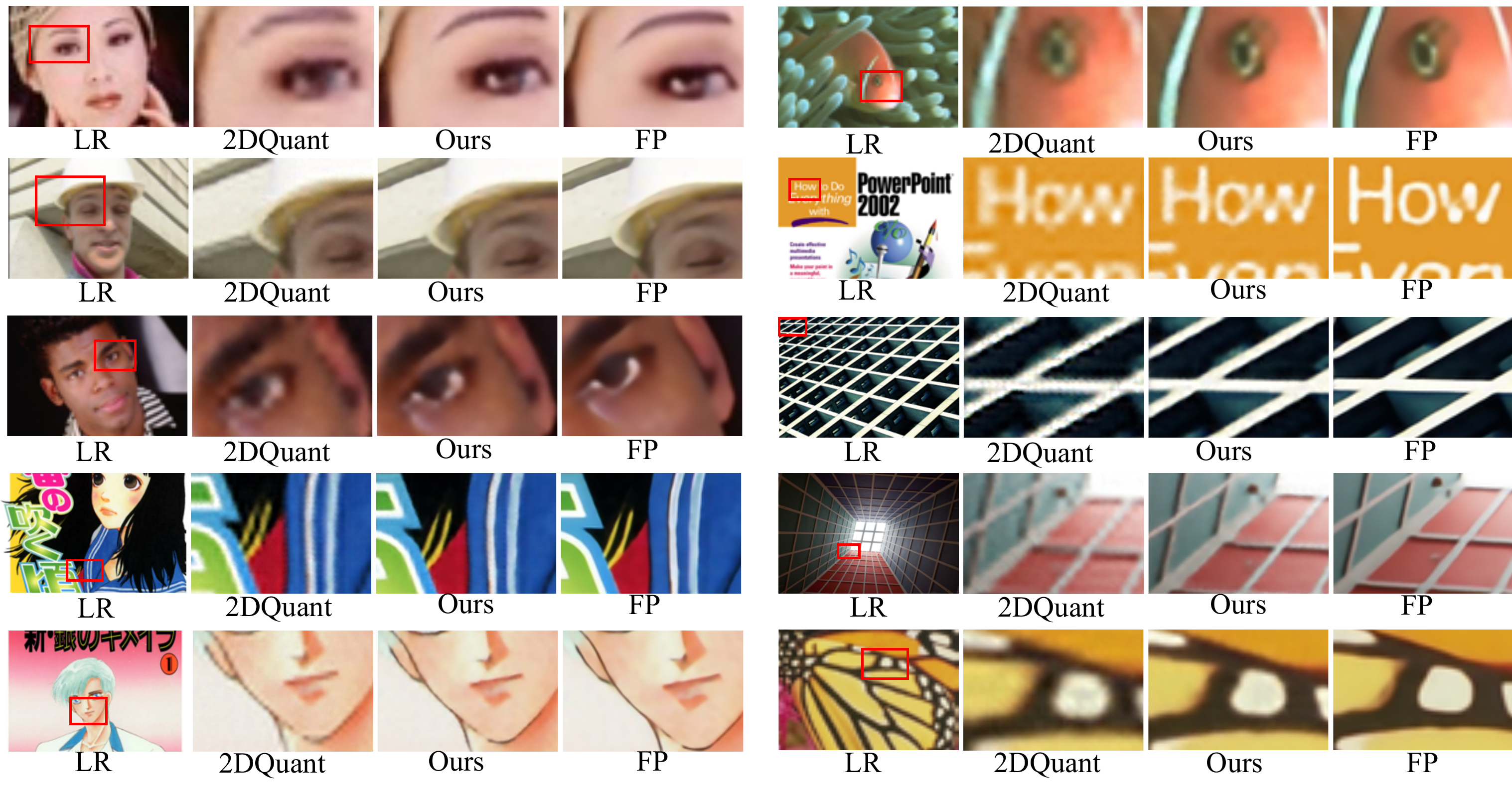}

\caption{Qualitative visual comparison for 2-bit $(\times 4)$ SR across all five benchmark datasets. LR denotes the low-resolution input. Reference outputs are from the FP SwinIR-light~\citep{liang2021swinir} and the 2-bit $(\times 4)$ 2DQuant baseline~\citep{liu20242dquant}.}
\vspace{-2mm}
\label{fig:Qualitative}
\end{figure*}

\Cref{fig:Qualitative} compares 2-bit $(\times 4)$ outputs of our model and 2DQuant's, from the same low-resolution input. Across natural images and manga illustrations our outputs are sharper, with less blur and graininess. Manga109 shows clearer linework, Urban100 preserves architectural lines that 2DQuant blurs, the Set14 text is more legible, and the B100 and Set5 faces recover sharper detail. Implementation details are in \Cref{sec:impdetails}.


\subsection{Ablation Studies}
The two components of the method are ablated separately. The Hadamard is ablated across the whole grid: the ``CompSRT w/o Had.'' rows of \Cref{tab:performance_comparison3,tab:performance_comparison1,tab:performance_comparison2} are the identical recipe with the transform replaced by the identity, and the rotation improves PSNR in all $45$ configurations (\Cref{sec:additional}). The additive correction $\alpha,\beta$ is ablated in \Cref{tab:scalar-ablation} by holding it inert; that ablation also varies the bound learning, and finds it the larger effect.

\noindent\textbf{Is the effect Hadamard-specific?} \Cref{tab:transform-ablation} re-runs both tests under four alternative orthogonal kernels: range reduction replicates under every one ($d_z\in[0.36,0.42]$ on weights, $[0.51,0.85]$ on activations, $p<10^{-6}$), while the $\varepsilon$-band increase appears only under Sylvester+padding and reverses under the unpadded kernels. \Cref{sec:transform-ablation} carries the same substitutions through end-to-end training.

\subsection{Model Complexity}\label{sec:complexity}
We report inference time on Set5 ($\times 4$, all bitwidths) in \Cref{tab:complexity}; full scale coverage in \Cref{tab:complexity-all}. The rotation is not free: the Hadamard adds $0.16$--$0.20$\,s of CUDA-synced wall-clock per image versus 2DQuant on an A100-80GB, in the FP32 dequantize\,$\to$\,Hadamard\,$\to$\,quantize transition (\Cref{sec:rt-breakdown}). This is the practical cost of the gains in \Cref{tab:performance_comparison3}, and it is substantial here: a lightweight backbone cannot amortise the transform the way a billion-parameter LLM can.

\noindent\textbf{Bits per parameter.} We quantize weights and activations to a single per-tensor $(u,l)$ pair, so the storage cost is $b$ bits per weight plus two FP32 scalars and the two correction terms $\alpha,\beta$ per tensor --- $4.007$ bits/parameter at $b{=}4$, $3.007$ at $b{=}3$, $2.007$ at $b{=}2$ for SwinIR-light. The Sylvester construction pads each reduction axis to the next power of two, raising the cost to $4.274$/$3.207$/$2.140$ bits/parameter; \Cref{sec:bitcharge} charges the rotation for that difference. We report no megabyte column: the on-disk checkpoint is dominated by recomputable buffers --- a $12.00$\,MiB FP32 \texttt{attn\_mask} and a $0.75$\,MiB int64 \texttt{relative\_position\_index} --- rather than by weights, so a raw MB figure understates the compression threefold and is not a meaningful comparison.

\begin{figure*}[t]
\centering
\begin{minipage}[c]{0.47\textwidth}
  \centering
  \small
  \setlength{\tabcolsep}{3pt}
  \renewcommand{\arraystretch}{1}
  \begin{tabular}{c|c|c|c}
    \hline
    Method & Bit & $s$ & $b$ \\
    \hline
    2DQuant & 4 & 0.076 & 4.007 \\
    CompSRT & 4 & 0.242 & 4.274 \\
    \hline
    2DQuant & 3 & 0.077 & 3.007 \\
    CompSRT & 3 & 0.243 & 3.207 \\
    \hline
    2DQuant & 2 & 0.079 & 2.007 \\
    CompSRT & 2 & 0.246 & 2.140 \\
    \hline
  \end{tabular}
  \captionof{table}{Inference time ($s$, sec; CUDA-synced median, 20 passes / 8 warmups) and bits-per-parameter ($b$) for $(\times 4)$ on A100-80GB, at every bitwidth; \Cref{tab:complexity-all} extends this to $\times 2$ and $\times 3$. The Hadamard costs $\sim$3$\times$ the wall-clock of an un-rotated quantizer of the same bitwidth, and the overhead is near-constant in absolute terms ($0.166$--$0.167$\,s) rather than proportional to the bit budget. Under the Sylvester construction, padding each reduction axis to the next power of two also charges $0.267$/$0.200$/$0.133$ extra bits/parameter at $b{=}4$/$3$/$2$. The overhead sits in one place: a fused \texttt{F.linear} against the precomputed Hadamard matrix is $\sim$48\% of the forward pass (\Cref{sec:rt-breakdown}).}
  \label{tab:complexity}
\end{minipage}
\hfill
\begin{minipage}[c]{0.5\textwidth}
  \centering
  \small
  \setlength{\tabcolsep}{4pt}
  \renewcommand{\arraystretch}{1}
  \begin{tabular}{c|cc|cc}
    \hline
    \multirow{2}{*}{{Scale, Bit}}  & \multicolumn{2}{c|}{{PSNR}} & \multicolumn{2}{c}{{SSIM}} \\
     & $p$ & $d_z$ & $p$ & $d_z$ \\\hline
    $\times$2, 4 & 0.063 & 2.45 & 0.063 & --- \\\hline
    $\times$2, 3 & 0.031 & 2.08 & 0.031 & 1.46 \\\hline
    $\times$2, 2 & 0.031 & 1.60 & 0.031 & 1.71 \\\hline
    $\times$3, 4 & 0.031 & 2.08 & 0.031 & 1.58 \\\hline
    $\times$3, 3 & 0.031 & 2.18 & 0.031 & 1.92 \\\hline
    $\times$3, 2 & 0.031 & 1.71 & 0.031 & 3.02 \\\hline
    $\times$4, 4 & 0.031 & 2.76 & 0.031 & 5.66 \\\hline
    $\times$4, 3 & 0.031 & 1.94 & 0.031 & 3.89 \\\hline
    $\times$4, 2 & 0.031 & 2.16 & 0.031 & 3.37 \\\hline
  \end{tabular}
  \captionof{table}{One-sided Wilcoxon signed-rank tests (CompSRT $>$ 2DQuant) and paired effect sizes (Cohen's $d_z$) over the $N{=}5$ benchmarks, at the precision printed in \Cref{tab:performance_comparison3,tab:performance_comparison1,tab:performance_comparison2}. Eight of nine cells attain $p=0.031$, the minimum possible at $N{=}5$. The exception is $\times$2, 4-bit, where the Urban100 tie leaves $N{=}4$ (minimum attainable $p$ of $0.063$); all four remaining differences are positive in both metrics, and its SSIM $d_z$ is undefined because they are identical ($+0.001$). Differences are raw, at equal nominal bitwidth; \Cref{sec:bitcharge} repeats the test with the Sylvester padding charged.\label{tab:binom_combined}}
\end{minipage}
\vspace{-4mm}
\end{figure*}
\par\vspace{-1mm}
{\centering
\small
\setlength{\tabcolsep}{6pt}
\renewcommand{\arraystretch}{1}
\resizebox{\textwidth}{!}{%
\begin{tabular}{l|cc|cc|cc|cc}
\hline
\multirow{2}{*}{Orthogonal transform}
& \multicolumn{4}{c|}{Range $\Delta_r>0$}
& \multicolumn{4}{c}{$\varepsilon$-band $\Delta_p>0$} \\
\cline{2-9}
& \multicolumn{2}{c|}{Weights} & \multicolumn{2}{c|}{Acts}
& \multicolumn{2}{c|}{Weights} & \multicolumn{2}{c}{Acts} \\
& $d_z$ & $p$ & $d_z$ & $p$ & $d_z$ & $p$ & $d_z$ & $p$ \\
\hline
Sylvester Hadamard (+padding) & $+0.43$ & $8.7\mathrm{e}{-13}$ & $+0.95$ & $2.0\mathrm{e}{-35}$
                              & $+2.28$ & $4.0\mathrm{e}{-25}$ & $+0.61$ & $1.0\mathrm{e}{-4}$ \\
QuIP\#-Paley Hadamard         & $+0.38$ & $2.9\mathrm{e}{-7}$  & $+0.85$ & $2.4\mathrm{e}{-30}$
                              & $-0.38$ & $1$                  & $-0.16$ & $1$ \\
DCT-II                        & $+0.36$ & $4.2\mathrm{e}{-7}$  & $+0.53$ & $9.4\mathrm{e}{-16}$
                              & $-0.36$ & $1$                  & $-0.18$ & $1$ \\
Random orthogonal (seed 0)    & $+0.42$ & $1.6\mathrm{e}{-10}$ & $+0.67$ & $1.0\mathrm{e}{-21}$
                              & $-0.60$ & $1$                  & $-0.30$ & $1$ \\
Random orthogonal (seed 1)    & $+0.37$ & $2.5\mathrm{e}{-7}$  & $+0.51$ & $1.7\mathrm{e}{-15}$
                              & $-0.60$ & $1$                  & $-0.28$ & $1$ \\
\hline
\end{tabular}}
\captionof{table}{Range and $\varepsilon$-band Wilcoxon tests under five orthogonal transforms on the same calibration tensors ($N{=}144$ weights, $N{=}240$ activations); $d_z$ is the paired effect size, $p$ one-sided; per-transform detail is in \Cref{sec:transform-ablation}.}
\label{tab:transform-ablation}
\par}
\vspace{-1.5mm}
\section{Conclusion}
\vspace{-1mm}
We asked which property of the post-rotation distribution drives the gain from rotation, and answered it with a paired decomposition on an SR transformer. Range reduction is the kernel-independent mechanism, replicating under every orthogonal kernel we tried and on EDSR conv weights ($d_z{=}0.99$); the $\varepsilon$-band effect is a Sylvester-padding artifact. The end-task counterpart agrees: over a grid of three scales, three bitwidths and five benchmarks, the rotation improves PSNR in all $45$ configurations against an identical un-rotated quantizer, by more as the budget tightens. CompSRT improves on 2DQuant in $44$ of $45$ at equal nominal bitwidth, an edge that survives the padding charge at 3 and 2 bits but not at 4. \textbf{Limitations.} CondiQuant remains ahead at every operating point, by a mean of $0.12$\,dB at 4 bits widening to $0.65$\,dB at 2 bits; combining the two approaches is the obvious next step. The FP32 Hadamard transition roughly triples inference wall-clock and the padding costs $0.27$ bits/parameter, so the recipe needs a fused low-bit kernel. CompSRT on EDSR trails CNN-tuned PTQ by $\sim$$5.6$\,dB: the mechanism transfers across backbones, the recipe does not. Every end-task number is a single run at one seed, so our smallest margins are not separable from run-to-run noise.

\clearpage
\bibliographystyle{plainnat}
\bibliography{main}
\clearpage
\section{Supplementary Materials} 
\subsection{$\epsilon$-sensitivity of analysis}\label{sec:epssensitive}
To see whether we still observe a statistically significant increase in the proportion of values within $[-\varepsilon,\varepsilon]$ as $\varepsilon$ varies, we repeated our analysis with $\varepsilon \in [0.01,1]$ in increments of $0.1$, again testing $\mathrm{median}(\Delta_p)>0$. As $\varepsilon$ increases, $p$-values rise and effect sizes shrink, consistent with many tensors being distributed around $0$ with small standard deviation. However, most $p$-values remain significant, indicating the analysis is robust to $\varepsilon$ when sufficiently small.

\begin{table}[h]
    \centering
    \small
    \setlength{\tabcolsep}{6pt}
    \renewcommand{\arraystretch}{1.1}
    \begin{tabular}{c|cc|cc}
      \hline
      \multirow{2}{*}{$\epsilon$}
      & \multicolumn{2}{c|}{Weights}
      & \multicolumn{2}{c}{Activations} \\
      & $p$ & $d_z$ & $p$ & $d_z$ \\
      \hline
      0.01 & $1.1\mathrm{e}{-25}$ & 8.54 & $2.0\mathrm{e}{-8}$ & 0.72 \\
      0.1  & $5.0\mathrm{e}{-22}$ & 1.38 & $6.0\mathrm{e}{-4}$ & 0.57 \\
      0.2  & $2.5\mathrm{e}{-21}$ & 1.00 & $0.22$              & 0.25 \\
      0.3  & $1.2\mathrm{e}{-21}$ & 0.98 & $0.15$              & 0.33 \\
      0.4  & $2.8\mathrm{e}{-20}$ & 0.80 & $0.08$              & 0.33 \\
      0.5  & $5.6\mathrm{e}{-13}$ & 0.43 & $0.01$              & 0.29 \\
      0.6  & $2.5\mathrm{e}{-7}$  & 0.26 & $0.01$              & 0.24 \\
      0.7  & $1.7\mathrm{e}{-6}$  & 0.23 & $4\mathrm{e}{-3}$   & 0.20 \\
      0.8  & $9.9\mathrm{e}{-5}$  & 0.20 & $4\mathrm{e}{-3}$   & 0.18 \\
      0.9  & $0.001$              & 0.18 & $4\mathrm{e}{-3}$   & 0.16 \\
      1.0  & $0.01$               & 0.17 & $0.01$              & 0.15 \\
      \hline
    \end{tabular}
    \caption{Sensitivity of the $\varepsilon$-band test to the choice of $\varepsilon$. As $\varepsilon$ grows, $p$-values increase and $d_z$ decreases, but most remain significant at $\alpha{=}0.05$, indicating the analysis is robust to $\varepsilon$ in the small-band regime that we report in the main paper.}
    \label{tab:epsilonsens}
\end{table}

\begin{table}[h]
\centering
\small
\setlength{\tabcolsep}{4pt}
\renewcommand{\arraystretch}{1.05}
\begin{tabular}{l|c|c|c|c|c}
\hline
Condition $(\times 2$, 2-bit$)$ & Bounds $u,l$ & $\alpha,\beta$ & Had. & Set5 & Mean \\
\hline
Calibrated bounds held fixed        & fixed   & free  & ---  & 35.74 / 0.9488 & 32.08 \\ \hline
No rotation, correction inert       & learned & inert & ---  & 36.27 / 0.9508 & 32.64 \\ \hline
No rotation (CompSRT w/o Had.)      & learned & free  & ---  & 36.22 / 0.9509 & 32.63 \\ \hline
Rotation, correction inert          & learned & inert & WA   & 36.62 / 0.9526 & 33.07 \\ \hline
Full recipe (CompSRT)               & learned & free  & WA   & 36.47 / 0.9517 & 32.91 \\
\hline
\end{tabular}
\caption{Decomposition ablation at $\times 2$, 2-bit. ``Set5'' is PSNR\,/\,SSIM; ``Mean'' is mean PSNR over all five benchmarks, which we use for the comparisons below because it is less sensitive to a single test set. ``Inert'' holds $\alpha,\beta$ at zero; all conditions are otherwise identical, including the calibrated bound search, the finetuning schedule, and the bit budget. Three effects separate cleanly, in decreasing order of size. (i)~\emph{Learning the bounds dominates}: holding $u,l$ at their calibrated values costs $0.55$\,dB. (ii)~\emph{The rotation is worth $0.28$--$0.43$\,dB} and helps under every treatment of the correction. (iii)~\emph{The additive correction $\alpha,\beta$ does not improve accuracy}: it is indistinguishable from zero without the rotation ($-0.00$\,dB) and costs $0.15$\,dB with it, the latter negative on all five benchmarks. An independent route to the same contrast --- constraining $\alpha$ to be non-negative, which drives it to zero, rather than freezing it --- reproduces this: $32.71$ and $33.05$ mean PSNR for the two inert rows. The spread between the two routes ($0.07$\,dB unrotated, $0.01$\,dB rotated) bounds the precision of these comparisons, so we do not read the unrotated $\alpha,\beta$ effect as distinguishable from zero.}
\label{tab:scalar-ablation}
\end{table}

\begin{table}[h]
\centering
\small
\setlength{\tabcolsep}{5pt}
\renewcommand{\arraystretch}{1.05}
\resizebox{\textwidth}{!}{%
\begin{tabular}{l|l|c|c|c}
\hline
Variant & Modification vs.\ CompSRT & PSNR & SSIM & $\Delta$ PSNR \\
\hline
CompSRT (reference) & ---                                                              & 31.98 & 0.890 & ---       \\ \hline
DCT-CompSRT          & DCT-II in place of Hadamard                                      & 31.92 & 0.889 & $-0.06$    \\ \hline
RandOrth-CompSRT     & random orthogonal matrix in place of Hadamard                    & 31.91 & 0.889 & $-0.07$    \\ \hline
CompSRT-LSQ          & $\alpha,\beta$ replaced by learnable $S$ (LSQ-style)              & 32.00 & 0.890 & $+0.02$    \\
\hline
\end{tabular}}
\caption{End-to-end methodology ablations on Set5 ($\times 4$, 4-bit), each starting from the same calibration data and trained for up to 4000 iterations with the same hyperparameters as the headline CompSRT recipe. (i)~Substituting an orthonormal DCT-II for the Hadamard costs $0.06$\,dB, and a random orthogonal matrix $0.07$\,dB; both remain well above the un-rotated variant at $31.74$\,dB (\Cref{tab:performance_comparison3}). \emph{These rows are not at a matched bit budget.} Both substitutes are built at the tensor's own width and so run at $b{=}4.007$, while the Sylvester reference pads and runs at $b{=}4.274$ --- it is handed $0.267$ more bits per parameter than the rows it leads by $0.06$--$0.07$\,dB, and this arm's own ladder prices those bits at $\approx0.15$\,dB on Set5 (\Cref{sec:bitcharge}). Charged for its padding the Hadamard is at best level with the unpadded transforms, which is the reading consistent with \Cref{tab:transform-ablation}. Either way the choice of kernel is a runtime decision, not a quantization-quality one. (ii)~Replacing the additive correction $\alpha,\beta$ with LSQ-style direct step-size learning changes the result by $+0.02$\,dB --- within the precision of these comparisons (\Cref{tab:scalar-ablation}). How the step is parameterized is therefore not what determines accuracy here, which is consistent with the correction itself contributing no gain.}
\label{tab:methodology-ablations}
\end{table}

\subsection{Scalar Ablation Study} \label{sec:scalarablation}
To separate the contributions of bound learning, the additive correction $\alpha,\beta$, and the rotation, we ablate all three at $\times 2$, 2-bit. \Cref{tab:scalar-ablation} reports five conditions sharing one fake-quantization architecture, calibration, schedule and bit budget; we compare on the mean over the five benchmarks rather than on Set5 alone.

The three factors are strongly unequal. Learning the clipping bounds is the largest single effect: holding $u,l$ at their calibrated values costs $0.55$\,dB, more than the rotation contributes. The rotation is next, worth $0.28$\,dB in the full recipe and $0.43$\,dB when the correction is inert --- it helps under every treatment of $\alpha,\beta$, which is the form of the claim that matters for \Cref{sec:hadamard}. The additive correction is the smallest and does not improve accuracy: without the rotation it is indistinguishable from zero, and with the rotation it costs $0.15$\,dB, negative on all five benchmarks.

That the correction is not merely inert but mildly harmful in combination deserves comment, since the two are nominally independent. The correction acts as a learned tightening of the calibrated range (\Cref{sec:related}), and it tightens by about the same fraction with or without the rotation. What differs is the distribution it tightens. The calibrated bounds are set by the extremes of the tensor; on an un-rotated, heavy-tailed tensor those extremes are fixed by rare outliers, so discarding the outer fifth of the range removes little mass. A rotation mixes each coefficient into a sum over its row and drives the tensor toward a Gaussian, spreading mass across the range, so the same proportional tightening discards more of it. We report this as the reading most consistent with the measurements rather than as a verified mechanism: we measured the learned spans and the end-task effect, not the clipped mass.

The same contrast at $\times 4$, 4-bit is smaller and in the same direction --- the correction costs $0.02$\,dB unrotated and $0.05$\,dB rotated --- which is what the learned spans predict, since at 4-bit the median span is $0.93$ of the calibrated range for weights and $0.96$ for activations, against $0.71$ and $0.82$ at 2-bit. The correction has less room to act when there are more levels to spend.

\subsection{Where does the rotation act?}\label{sec:rotation-scope}
The recipe rotates both operands. Applying the transform to only one side separates their contributions; the inverse is applied on the same side, so each condition remains a faithful quantizer of the same network. At $\times 2$, 2-bit the two sides behave oppositely. Rotating activations alone is worth $+0.36$\,dB over no rotation, positive on all five benchmarks. Rotating weights alone \emph{costs} $0.22$\,dB, negative on all five. The full recipe sits at $+0.28$\,dB, between the two and below the activation-only condition.

The benefit is therefore an activation-side effect. This is consistent with where the range problem lies: weights are static and their calibrated bounds already fit them well, whereas activations vary with the input and are the operand whose dynamic range the calibrated bounds must cover conservatively. It also means the recipe is not the best configuration in this family --- rotating activations alone is $0.08$\,dB better at this operating point. We report the two-sided recipe because it is what the rest of the paper measures, and note that a one-sided variant would also halve the transform's runtime cost.

\subsection{Verifying the realised bitwidth}\label{sec:bitaudit}
A quantizer that learns any part of its step size can, if the integer code is left unbounded, represent more levels than its nominal bitwidth advertises, in which case a reported ``$b$-bit'' number is not comparable with a baseline at the same $b$. The clamp of \Cref{eq:codeclamp} rules this out by construction, but we verify it empirically rather than assert it, since it is cheap to check and expensive to get wrong.

We instrument every quantizer in each of the $18$ trained models (three scales $\times$ three bitwidths $\times$ \{CompSRT, CompSRT w/o Had.\}) and record, over a full evaluation pass, the bitwidth each quantizer actually holds and the number of distinct values it emits. Across all $18$ models and all $384$ quantizers per model ($144$ weight, $240$ activation), the recorded bitwidth equals the nominal $b$ in every case, and no quantizer is a pass-through. For the un-rotated models the number of distinct emitted values is exactly $2^b$ --- $16$, $8$ and $4$ at $b={4,3,2}$ --- confirming the clamp binds.

One caveat affects how such an audit should be read. For the rotated models the same probe reports far more than $2^b$ distinct values (medians in the hundreds to low thousands). That is not a violation of the bit budget: the inverse rotation is applied inside the quantizer's forward pass, so its outputs are linear combinations of many $b$-bit codes rather than the codes themselves. Counting distinct output values is a valid bitwidth probe only for transform-free quantizers and reports a spuriously high figure for any rotation-based method; the recorded per-quantizer bitwidth stays meaningful in both cases. We note it because the distinct-value count is the more obvious check to reach for.

\subsection{Implementation details}\label{sec:impdetails}
We build our work on top of 2DQuant's open sourced repository \cite{liu20242dquant}, and use SwinIR-light~\citep{liang2021swinir} as the model backbone of our method. For the Hadamard transform and statistical tests, we use SciPy \cite{2020SciPy-NMeth}. For finetuning, we use the Adam \cite{kinga2015method} optimizer with a learning rate of $1*10^{-2}$ and betas set to (0.9, 0.999). We clip all gradient values to between $[-1,1]$ and we finetune for at most 4000 iterations, or until we reach a nan gradient which we handle by safely exiting. Validation runs every $100$ iterations on Set5 and we select the highest-PSNR checkpoint for evaluation on all five benchmarks; the same protocol is applied to every arm of every ablation, so no comparison in this paper comes from mixing a best-validation checkpoint with a final iterate. For measuring the time and space complexity of our model, we calculate inference time in seconds and storage in bits per parameter. Every configuration reported in this paper is a single finetuning run at one seed; we did not repeat runs, so no result here carries an error bar (\Cref{sec:bitcharge}). Our code is written with PyTorch \cite{paszke2019pytorch}; training and finetuning run for at most 4 hours on one NVIDIA RTX~6000 48G GPU. The inference-time numbers reported in \Cref{tab:complexity}, \Cref{tab:complexity-all}, and \Cref{tab:rt-breakdown} are measured separately on a single NVIDIA A100-80GB to provide a deployment-realistic timing benchmark; all use CUDA-synchronized timing. The open source code for this paper along with instructions for reproducibility can be found \href{https://anonymous.4open.science/r/CompSRT-439F}{here}.
\subsection{Additional Results} \label{sec:additional} 
We report the SwinIR-light $\times 2$ and $\times 3$ models across all quantization bitwidths in \Cref{tab:performance_comparison1} and \Cref{tab:performance_comparison2}; together with \Cref{tab:performance_comparison3} these cover the full $3\times3\times5$ grid. Two patterns hold across the whole grid. First, the Hadamard improves PSNR in every one of the $45$ configurations relative to the un-rotated variant at the same nominal bitwidth, with the smallest gain $+0.08$\,dB (B100, 4-bit $\times 2$) and the largest $+0.62$\,dB (Manga109, at both 3-bit $\times 3$ and 2-bit $\times 4$); the mean gain per (scale, bit) cell ranges from $+0.18$ to $+0.34$\,dB and is generally larger at tighter budgets, monotonically so at $\times 2$ and $\times 4$ (the one exception is $\times 3$, where the 3-bit mean of $+0.34$\,dB exceeds the 2-bit mean of $+0.25$\,dB). The rotated arm carries the Sylvester zero-padding, $0.267$/$0.200$/$0.133$ more bits per parameter at $4$/$3$/$2$ bits; charged for that at this model's own bit-rate, the gain narrows to $+0.02$--$0.20$\,dB and remains positive in all nine (scale, bit) cells (\Cref{sec:bitcharge}). Second, CompSRT improves on 2DQuant in $44$ of the $45$ configurations, the exception being a tie on Urban100 at 4-bit $\times 2$; charged for the padding that advantage survives at 3 and 2 bits but not at 4 (\Cref{sec:bitcharge}). CondiQuant is ahead of CompSRT in all $45$; the deficit is smallest at 4 bits (mean $0.12$\,dB) and grows to a mean of $0.65$\,dB at 2 bits, its largest value being $1.40$\,dB on Manga109 at 2-bit $\times 3$. The rotation is therefore a reliable improvement over an un-rotated quantizer of the same bitwidth, but distribution shaping on its own does not overtake conditioning-based PTQ in the low-bit regime.
\begin{table*}[t]
\centering
\small
\setlength{\tabcolsep}{3pt}
\renewcommand{\arraystretch}{1}
\begin{tabularx}{\textwidth}{l|c|XX|XX|XX|XX|XX}
\hline

\multirowcell{2}[-0.5ex][c]{{Method ($\times$2)}} &
\multirowcell{2}[-0.5ex][c]{{Bit}} &

\multicolumn{2}{c|}{{Set5 }} & 
\multicolumn{2}{c|}{{Set14 }} & 
\multicolumn{2}{c|}{{B100}} & 
\multicolumn{2}{c|}{{Urban100}} & 
\multicolumn{2}{c}{{Manga109}} 
\\
& & {PSNR} & {SSIM} & {PSNR} & {SSIM} & {PSNR} & {SSIM} & {PSNR} & {SSIM} & {PSNR} & {SSIM} \\
\hline
SwinIR-light & 32 & 38.15 & 0.961 & 33.86 & 0.921 & 32.31 & 0.901 & 32.76 & 0.934 & 39.11 & 0.978 \\
Bicubic & 32 & 32.25 & 0.912 & 29.25 & 0.841 & 28.68 & 0.810 & 25.96 & 0.809 & 29.17 & 0.913 \\
\noalign{\vskip 1pt}\hline\noalign{\vskip 1pt}
PTQ4ViT & 4 & 37.43 & 0.957 & 33.19 & 0.914 & 31.84 & 0.895 & 31.54 & 0.921 & 37.59 & 0.974 \\
NoisyQuant & 4 & 37.50 & 0.957 & 33.06 & 0.910 & 31.73 & 0.894 & 31.31 & 0.918 & 37.47 & 0.972 \\
2DQuant & 4 & 37.87 & 0.959 & 33.41 & 0.916 & 32.02 & 0.897 & 31.84 & 0.925 & 38.31 & 0.976 \\
CondiQuant  & 4 & \textbf{38.03} & \textbf{0.961} & \textbf{33.50} & \textbf{0.918} & \textbf{32.16} & \textbf{0.899} & \textbf{32.03} & \textbf{0.928} & \textbf{38.57} & \textbf{0.977} \\
CompSRT w/o Had. & 4 & 37.79 & 0.959 & 33.29 & 0.915 & 31.98 & 0.897 & 31.50 & 0.922 & 38.07 & 0.975 \\
CompSRT & 4 & 37.89 & 0.960 & 33.47 & 0.917 & 32.06 & 0.898 & 31.84 & 0.926 & 38.35 & 0.976 \\
\noalign{\vskip 1pt}\hline\noalign{\vskip 1pt}
PTQ4ViT & 3 & 36.49 & 0.951 & 32.49 & 0.905 & 31.27 & 0.885 & 30.16 & 0.903 & 36.41 & 0.967 \\
NoisyQuant & 3 & 35.32 & 0.933 & 31.88 & 0.891 & 30.73 & 0.871 & 29.28 & 0.884 & 35.30 & 0.954 \\
2DQuant & 3 & 37.32 & 0.957 & 32.85 & 0.911 & 31.60 & 0.891 & 30.45 & 0.909 & 37.24 & 0.972 \\
CondiQuant & 3 & \textbf{37.77} & \textbf{0.959} & \textbf{33.21} & \textbf{0.915} & \textbf{31.94} & \textbf{0.897} & \textbf{31.18} & \textbf{0.920} & \textbf{38.01} & \textbf{0.976} \\
CompSRT w/o Had. & 3 & 37.30 & 0.957 & 32.85 & 0.911 & 31.63 & 0.892 & 30.50 & 0.910 & 37.28 & 0.972 \\
CompSRT & 3 & 37.48 & 0.958 & 33.06 & 0.913 & 31.77 & 0.894 & 30.90 & 0.915 & 37.61 & 0.974 \\
\noalign{\vskip 1pt}\hline\noalign{\vskip 1pt}
PTQ4ViT & 2 & 33.25 & 0.892 & 30.22 & 0.840 & 29.21 & 0.807 & 27.31 & 0.811 & 32.75 & 0.909 \\
NoisyQuant & 2 & 30.13 & 0.762 & 28.80 & 0.756 & 28.26 & 0.742 & 26.68 & 0.763 & 30.40 & 0.820 \\
2DQuant & 2 &  {36.00} &  {0.950} &  {31.98} &  {0.901} &  {30.91} &  {0.881} &  {28.62} &  {0.882} &  {34.40} &  {0.960} \\
CondiQuant & 2 & \textbf{37.15} & \textbf{0.957} & \textbf{32.74} & \textbf{0.910} & \textbf{31.55} & \textbf{0.891} & \textbf{29.96} & \textbf{0.905} & \textbf{36.63} & \textbf{0.971} \\
CompSRT w/o Had. & 2 & 36.22 & 0.951 & 32.14 & 0.903 & 31.02 & 0.883 & 28.92 & 0.886 & 34.87 & 0.963 \\
CompSRT & 2 & 36.47 & 0.952 & 32.30 & 0.904 & 31.14 & 0.884 & 29.17 & 0.890 & 35.48 & 0.964 \\
\end{tabularx}
\par\smallskip
\caption{Performance comparison for scale factor $(\times2)$ across bit widths; best in each column in \textbf{bold}. Baselines are as reported in their papers; CompSRT rows are measured here. ``CompSRT w/o Had.'' replaces the Hadamard with the identity at the same nominal bitwidth; the rotated arm additionally carries the Sylvester zero-padding, so it stores $0.267$/$0.200$/$0.133$ more bits per parameter at $4$/$3$/$2$ bits. CompSRT improves on the un-rotated variant in every cell and on 2DQuant in $14$ of $15$ (Urban100 at 4-bit ties at $31.84$\,dB); CondiQuant leads throughout by $0.03$--$1.15$\,dB.}
\label{tab:performance_comparison1}
\end{table*}
\begin{table*}[ht]
\centering
\small
\setlength{\tabcolsep}{3pt}
\renewcommand{\arraystretch}{1}
\begin{tabularx}{\textwidth}{l|c|XX|XX|XX|XX|XX}
\hline

\multirowcell{2}[-0.5ex][c]{{Method ($\times$3)}} &
\multirowcell{2}[-0.5ex][c]{{Bit}} &

\multicolumn{2}{c|}{{Set5 }} & 
\multicolumn{2}{c|}{{Set14 }} & 
\multicolumn{2}{c|}{{B100}} & 
\multicolumn{2}{c|}{{Urban100}} & 
\multicolumn{2}{c}{{Manga109}} 
\\
& & {PSNR} & {SSIM} & {PSNR} & {SSIM} & {PSNR} & {SSIM} & {PSNR} & {SSIM} & {PSNR} & {SSIM} \\
\hline
SwinIR-light & 32 & 34.63 & 0.929 & 30.54 & 0.846 & 29.20 & 0.808 & 28.66 & 0.862 & 33.99 & 0.948 \\
Bicubic & 32 & 29.54 & 0.852 & 27.04 & 0.755 & 26.78 & 0.719 & 24.00 & 0.714 & 26.16 & 0.838 \\
\noalign{\vskip 1pt}\hline\noalign{\vskip 1pt}
PTQ4ViT & 4 & 33.77 & 0.920 & 29.75 & 0.827 & 28.62 & 0.794 & 27.43 & 0.836 & 32.50 & 0.936 \\
NoisyQuant & 4 & 33.13 & 0.912 & 29.06 & 0.809 & 27.93 & 0.775 & 26.66 & 0.814 & 31.94 & 0.929 \\
2DQuant & 4 & 34.06 & 0.923 &  {30.12} &  {0.837} &  {28.89} & 0.799 & 27.69 & 0.841 & 32.88 & 0.939 \\
CondiQuant & 4 & \textbf{34.32} & \textbf{0.926} & \textbf{30.29} & \textbf{0.842} & \textbf{29.05} & \textbf{0.804} & \textbf{28.05} & \textbf{0.851} & \textbf{33.23} & \textbf{0.943} \\
CompSRT w/o Had. & 4 & 34.05 & 0.923 & 30.09 & 0.837 & 28.87 & 0.799 & 27.66 & 0.840 & 32.79 & 0.939 \\
CompSRT & 4 & 34.17 & 0.924 & 30.21 & 0.839 & 28.97 & 0.801 & 27.89 & 0.846 & 33.11 & 0.941 \\
\noalign{\vskip 1pt}\hline\noalign{\vskip 1pt}
PTQ4ViT & 3 & 32.75 & 0.903 & 29.14 & 0.811 & 28.06 & 0.771 & 26.43 & 0.801 & 31.20 & 0.913 \\
NoisyQuant & 3 & 30.78 & 0.851 & 27.94 & 0.762 & 26.98 & 0.715 & 25.43 & 0.748 & 29.64 & 0.879 \\
2DQuant & 3 &  {33.24} &  {0.914} &  {29.56} &  {0.826} &  {28.50} &  {0.787} &  {26.65} &  {0.812} &  {31.46} &  {0.924} \\
CondiQuant & 3 & \textbf{33.92} & \textbf{0.922} & \textbf{30.02} & \textbf{0.837} & \textbf{28.84} & \textbf{0.799} & \textbf{27.37} & \textbf{0.836} & \textbf{32.48} & \textbf{0.937} \\
CompSRT w/o Had. & 3 & 33.33 & 0.915 & 29.62 & 0.827 & 28.53 & 0.788 & 26.69 & 0.814 & 31.50 & 0.923 \\
CompSRT & 3 & 33.64 & 0.918 & 29.83 & 0.832 & 28.69 & 0.793 & 27.11 & 0.826 & 32.12 & 0.931 \\
\noalign{\vskip 1pt}\hline\noalign{\vskip 1pt}
PTQ4ViT & 2 & 29.96 & 0.790 & 27.36 & 0.703 & 26.74 & 0.659 & 24.56 & 0.646 & 27.37 & 0.739 \\
NoisyQuant & 2 & 27.53 & 0.664 & 25.77 & 0.595 & 25.37 & 0.561 & 23.59 & 0.574 & 26.03 & 0.663 \\
2DQuant & 2 &  {31.62} &  {0.889} &  {28.54} &  {0.804} &  {27.85} &  {0.768} &  {25.30} &  {0.769} &  {28.46} &  {0.881} \\
CondiQuant & 2 & \textbf{33.00} & \textbf{0.913} & \textbf{29.44} & \textbf{0.825} & \textbf{28.45} & \textbf{0.788} & \textbf{26.36} & \textbf{0.808} & \textbf{30.88} & \textbf{0.920} \\
CompSRT w/o Had. & 2 & 31.92 & 0.894 & 28.76 & 0.808 & 27.96 & 0.772 & 25.49 & 0.776 & 28.98 & 0.890 \\
CompSRT & 2 & 32.19 & 0.898 & 28.92 & 0.811 & 28.08 & 0.775 & 25.70 & 0.782 & 29.48 & 0.895 \\
\hline
\end{tabularx}
\caption{Performance comparison for scale factor $(\times3)$ across bit widths; best in each column in \textbf{bold}. Baselines are as reported in their papers; CompSRT rows are measured here. ``CompSRT w/o Had.'' replaces the Hadamard with the identity at the same nominal bitwidth; the rotated arm additionally carries the Sylvester zero-padding, so it stores $0.267$/$0.200$/$0.133$ more bits per parameter at $4$/$3$/$2$ bits. CompSRT improves on the un-rotated variant and on 2DQuant in all $15$ cells; CondiQuant leads throughout by $0.08$--$1.40$\,dB.}
\label{tab:performance_comparison2}
\end{table*}

\subsection{Significance testing procedure}\label{sec:results-significance}
For a fixed scale and bit width and for each metric $m\in\{\text{PSNR},\text{SSIM}\}$, we form per-dataset differences $\Delta_i = \text{score}_{\text{CompSRT},i}^{(m)}- \text{score}_{\text{2DQuant},i}^{(m)}$ for $i=1,\dots,N$ (with $N{=}5$ datasets), and test $H_0:\mathrm{median}(\Delta)=0$ against $H_1:\mathrm{median}(\Delta)>0$ with the paired Wilcoxon signed-rank test; exact ties ($\Delta_i=0$) are excluded. We report the one-sided $p$-value and Cohen's $d_z=\bar{\Delta}/s_\Delta$, where $\bar{\Delta}$ is the mean of the paired differences and $s_\Delta$ their sample standard deviation. The procedure runs independently for each (scale, bit, metric) configuration on dataset-level PSNR and SSIM averages, at the precision printed in the tables. At 4-bit $\times 2$ the Urban100 tie removes one pair, so $N{=}4$ and no result in that cell can reach $\alpha{=}0.05$ for either metric, even though all four remaining differences are positive. \Cref{sec:bitcharge} repeats the procedure on differences charged for the Sylvester padding.

\subsection{Charging the Sylvester padding}\label{sec:bitcharge}
Every end-task comparison in this paper is between arms of equal \emph{nominal} bitwidth, not equal storage. The rotated arm stores $4.274$/$3.207$/$2.140$ bits per parameter at $4$/$3$/$2$ bits, against $4.007$/$3.007$/$2.007$ for the un-rotated arm and for 2DQuant, so it is handed a surcharge of $0.267$/$0.200$/$0.133$ bits per parameter (\Cref{sec:complexity}). This section prices that surcharge and re-reads the comparisons with it charged. Everything below is derived from \Cref{tab:performance_comparison3,tab:performance_comparison1,tab:performance_comparison2} and can be recomputed from them.

\noindent\textbf{What a bit is worth.} We price bits on CompSRT's own ladder rather than assume a rate. For each scale and benchmark we take the secant between adjacent rungs: the $4.274\!\to\!3.207$ secant prices the 4-bit surcharge, and the $3.207\!\to\!2.140$ secant prices the 3- and 2-bit surcharges. Averaged over the five benchmarks these are $0.52$/$0.56$/$0.56$\,dB per bit on the upper segment at $\times 2$/$\times 3$/$\times 4$ and $1.17$/$1.32$/$1.09$ on the lower. PSNR is concave in $b$ --- the lower segment is roughly twice as steep --- so pricing bits \emph{above} $4.274$ at the upper secant charges more than the marginal bit is worth. Every net below is therefore a lower bound.

\noindent\textbf{The rotation survives the charge.} Charged, the rotation's mean advantage over the un-rotated arm is positive in all nine (scale, bit) cells, from $+0.02$\,dB ($\times 2$, 3-bit) to $+0.20$\,dB ($\times 4$, 2-bit), against raw means of $+0.18$ to $+0.34$\,dB.

\noindent\textbf{The comparison against 2DQuant does not, at 4 bits.} Charged, CompSRT's mean advantage over 2DQuant is positive in seven of the nine cells and negative in two, both at 4 bits: $-0.11$\,dB at $\times 2$ and $-0.01$\,dB at $\times 3$, with $\times 4$ marginal at $+0.01$\,dB. At 3 and 2 bits it stays positive ($+0.04$ to $+0.37$\,dB). Re-running the signed-rank test of \Cref{sec:results-significance} on the charged differences, significance at $\alpha{=}0.05$ holds in five of nine cells --- all three 2-bit cells, two of three 3-bit cells, and none of the 4-bit cells --- against eight of nine raw. The reading is that the rotation buys real accuracy at tight budgets, where a fraction of a bit is cheap relative to what the rotation recovers, and that at 4 bits CompSRT's edge over 2DQuant is bought with storage rather than with the transform.

Cell means are unaffected by whether bits are priced per benchmark or once per cell. Per-\emph{configuration} counts are not: charged, CompSRT beats 2DQuant in $33$ of $45$ configurations under per-benchmark pricing and $29$ of $45$ under a single per-cell rate, and the rotation beats the un-rotated arm in $41$ and $29$ respectively. We report cell means because they do not depend on that choice.

One limit applies to every end-task number in this paper, and it binds hardest here. Each (arm, scale, bitwidth) configuration is a \emph{single} finetuning run; we have not repeated any of them under different seeds, so we cannot bound run-to-run variance. The distributional results do not depend on this --- they are paired tests over $N{=}144$ and $N{=}240$ tensors from a fixed calibration pass --- but the charged end-task margins do. The smallest of them, $+0.01$--$0.02$\,dB, are far inside what a repeat could plausibly move, and the kernel-substitution rows of \Cref{tab:methodology-ablations} are single runs on one benchmark. We therefore read the charged results as showing which comparisons \emph{survive} the surcharge and which do not, not as measuring the size of what survives.

\subsection{Complexity for all Models} 
For scaling factors $\times 2$, $\times 3$, $\times 4$ we report the time for a single forward pass on an image from Set5, and the bits per parameter $b$, comparing with 2DQuant. The results are in \Cref{tab:complexity-all} and extend \Cref{tab:complexity} to all scales and bitwidths. The Hadamard adds roughly $0.16$--$0.20$\,s of CUDA-synchronized inference time per image across every configuration; the overhead is close to constant in absolute terms, so it is proportionally worst at $\times 4$, where the un-rotated forward pass is cheapest. The Sylvester padding is charged in $b$ throughout (\Cref{sec:bitcharge}). 
\begin{table}[h]
\centering
\small
\setlength{\tabcolsep}{5pt}
\renewcommand{\arraystretch}{1.05}
\begin{tabular}{l|c|c|c|c|c|c|c}
\hline
\multirow{2}{*}{Method} & \multirow{2}{*}{Bit}
& \multicolumn{2}{c|}{Scale $\times 2$}
& \multicolumn{2}{c|}{Scale $\times 3$}
& \multicolumn{2}{c}{Scale $\times 4$} \\
\cline{3-8}
& & $s$ & $b$ & $s$ & $b$ & $s$ & $b$ \\
\hline
2DQuant & 4 & 0.129 & 4.007 & 0.106 & 4.007 & 0.076 & 4.007 \\
CompSRT & 4 & 0.299 & 4.274 & 0.278 & 4.274 & 0.242 & 4.274 \\
\hline
2DQuant & 3 & 0.120 & 3.007 & 0.095 & 3.007 & 0.077 & 3.007 \\
CompSRT & 3 & 0.295 & 3.207 & 0.266 & 3.207 & 0.243 & 3.207 \\
\hline
2DQuant & 2 & 0.126 & 2.007 & 0.096 & 2.007 & 0.079 & 2.007 \\
CompSRT & 2 & 0.294 & 2.140 & 0.287 & 2.140 & 0.246 & 2.140 \\
\hline
\end{tabular}
\caption{Inference time in seconds per image ($s$, CUDA-synchronized median over 20 passes after 8 warmups on NVIDIA A100-80GB) and bits-per-parameter $b$ across scale factors $\times 2$, $\times 3$, $\times 4$ for bit widths 2--4, comparing with 2DQuant. $b$ charges the per-tensor quantization metadata for both methods and, for CompSRT, the Sylvester zero padding. The Hadamard transform adds approximately $0.16$--$0.20$\,s per forward pass across all configurations.}
\label{tab:complexity-all}
\end{table}

\subsection{Does the mechanism transfer to another architecture?}\label{sec:mambairv2}
The distributional analysis of \Cref{sec:hadamard} is a claim about what the Hadamard does to tensors, not about SwinIR-light specifically, so it should replicate on an unrelated backbone. We test it on MambaIRv2-light, an attentive state-space model~\citep{guo2024mambairv2}, by calibrating it and dumping pre/post-Hadamard tensor pairs from its attention, linear and batch-matmul layers. We report the distributional tests only; we do not report end-task PSNR for a quantized MambaIRv2-light.

We also re-run the range and $\varepsilon$-band tests of \Cref{tab:wtest-combined} on the freshly calibrated MambaIRv2-light pre/post-Hadamard tensors ($N{=}260$ weight pairs, $N{=}220$ activation pairs). Range reduction replicates with $d_z{=}0.40$ on weights ($p{=}7.7\!\times\!10^{-11}$) and $d_z{=}0.67$ on activations ($p{=}1.3\!\times\!10^{-36}$) --- consistent with our SwinIR-light effect sizes ($0.43/0.95$) and confirming the mechanism transfers to a different transformer family. The $\varepsilon$-band proportion is not significantly increased under the QuIP\#-Paley kernel here either ($d_z{=}-0.23$ for weights, $d_z{=}-0.65$ for activations), matching the construction-dependent pattern observed for SwinIR-light.

\subsection{Sylvester padding vs.\ QuIP\#-style Paley construction}\label{sec:padded-defense}
The Sylvester construction requires zero-padding to the next power of two, which under the SwinIR-light embedding dims $60, 120, 180$ pads to $64, 128, 256$. The QuIP\#-style Paley/Williamson construction~\citep{tseng2024quip} provides Hadamard matrices for orders $\{12, 20, 28, 36, 52, 60, 108, 116, 124, 140, 156, 172\}$, which combined with Sylvester butterflies covers $K \cdot 2^k$ for any $K$ in this set --- so $60$ and $120$ require no padding and $180$ pads to $192=12{\cdot}16$ (a $4\%$ pad rather than $42\%$). We re-run the range and $\varepsilon$-band Wilcoxon tests on the same set of $N{=}144$ weight tensors and $N{=}240$ activation tensors under both kernels; they are the first two rows of \Cref{tab:transform-ablation}. The range Wilcoxon test is robust to the kernel choice with similar effect sizes ($d_z=0.43$ vs.\ $0.38$ for weights, $0.95$ vs.\ $0.85$ for activations; both $p<10^{-6}$). The $\varepsilon$-band test is not: under the Paley kernel that does not pad $60$ or $120$, the per-tensor $\varepsilon$-band fraction is roughly unchanged or slightly decreased ($d_z=-0.38$ for weights, $d_z=-0.16$ for activations), indicating that the Sylvester $d_z=2.28$ reported in \Cref{tab:wtest-combined} is partly an artifact of comparing a $60$-dim distribution against a $64$-dim padded transform with extra near-zero entries. We treat range reduction as the primary mechanism for the Hadamard's quantization benefit in SR transformers, with normality improvement as a secondary mechanism that also replicates under both kernels.

\subsection{Inference-time breakdown}\label{sec:rt-breakdown}
We profile a single forward pass of CompSRT 4-bit $\times 4$ on a Set5 image and report how much of the wall-clock time is spent in the Hadamard transforms versus the rest of the pipeline, under three Hadamard implementations. Hadamard call sites are wrapped with CUDA-synchronized timers; the remainder is quantization/dequantization arithmetic, attention/MLP matmuls, normalization, and the upsample head. We additionally compare against an architecturally identical baseline in which the Hadamard call is replaced by an identity. Results are in \Cref{tab:rt-breakdown}. The fastest path is the \emph{Sylvester} implementation: a single fused \texttt{F.linear} with the precomputed Hadamard matrix takes 191\,ms ($\sim$48\% of total). The pure-PyTorch \emph{QuIP\#-Paley} port is much slower (15.0\,s) because the recursive butterfly fans out into many small Python-level kernel launches. Replacing the butterfly with Tri Dao's \texttt{fast\_hadamard\_transform} CUDA kernel~\citep{dao2024fasthadamard} brings the Paley path back to 807\,ms ($\sim$82\% of total): a $\sim$$19\times$ speedup over the unfused port. The fused Paley kernel does \emph{not} match Sylvester here, however, because Tri Dao's kernel is optimized for the very large pow2 dims that appear in LLMs (1024--16384) rather than the 60/120/192 dims that appear in SwinIR-light, where the kernel-launch overhead dominates the actual transform work, and the Paley path also adds a $K$-block \texttt{matmul} on top of the butterfly. The take-away is that the choice between Sylvester+padding and QuIP\#-Paley is a trade-off in our setting: Sylvester is fastest in wall-clock; QuIP\#-Paley with Tri Dao remains slower than 2DQuant ($\sim$76\,ms for 4-bit $\times 4$ on the same A100, \Cref{tab:complexity-all}) but the gap is dominated by the FP32 Hadamard transition rather than the kernel itself, and we use this construction for the $\varepsilon$-band kernel-comparison analysis in \Cref{sec:padded-defense}. The Hadamard overhead is more pronounced in small SR transformers than in LLMs because the surrounding matmuls are too small to amortize the FP32 transition cost.

\begin{table}[h]
\centering
\small
\setlength{\tabcolsep}{5pt}
\renewcommand{\arraystretch}{1.05}
\resizebox{\textwidth}{!}{%
\begin{tabular}{l|c|c|c|c}
\hline
\multirow{2}{*}{Component (per image, ms)} & \multicolumn{3}{c|}{Hadamard kernel} & \multirow{2}{*}{Identity} \\
\cline{2-4}
& Sylvester & QuIP\#-Paley (PyTorch) & QuIP\#-Paley + Tri Dao~\citep{dao2024fasthadamard} & \\
\hline
Hadamard transforms (fwd+inv, 768 calls)      & 191.4 (48.3\%) & 14994.9 (98.3\%) & 806.6 (81.9\%) & --- \\
Quantize/dequant, matmul, attention, upsample & 205.0 (51.7\%) & 255.5 (1.7\%)    & 178.0 (18.1\%) & 127.9 \\
\hline
\textbf{Total wall-clock per image}           & \textbf{396.4} & \textbf{15250.4} & \textbf{984.5} & \textbf{127.9} \\
\hline
Marginal overhead vs.\ identity baseline      & +268.5 (+210\%) & +15122.5 (+11824\%) & +856.6 (+670\%) & --- \\
\hline
\end{tabular}}
\caption{Inference-time breakdown for CompSRT 4-bit $\times 4$ on a Set5 image (NVIDIA A100-80GB; mean of 10 forward passes after 3 warmup runs). \textbf{All values are in milliseconds (ms).} Sylvester uses a single fused \texttt{F.linear} with a pow-of-two-padded Hadamard matrix. QuIP\#-Paley (PyTorch) is our pure-PyTorch port of the recursive butterfly + precomputed Paley tables of orders $\{12, 20, 60\}$; QuIP\#-Paley + Tri Dao replaces the butterfly with the fused \texttt{fast\_hadamard\_transform} CUDA kernel, which restores most of the per-call speed but does not match Sylvester here because the kernel is optimized for the much larger pow2 dims used in LLMs ($\geq 1024$) rather than the 60/120/192 dims in SwinIR-light, where launch overhead dominates and the $K$-block \texttt{matmul} on top of the butterfly costs extra. Wall-clock numbers are larger than \Cref{tab:complexity}/\Cref{tab:complexity-all} on the same hardware because this breakdown wraps each of 768 Hadamard call sites with an explicit \texttt{cuda.synchronize()} barrier; that profiling instrumentation adds $\sim$$150$\,ms of pure host--GPU sync overhead per image that is not present in the deployment-realistic timing reported in \Cref{tab:complexity}/\Cref{tab:complexity-all}.}
\label{tab:rt-breakdown}
\end{table}

\subsection{Is range reduction Hadamard-specific?}\label{sec:transform-ablation}
We run the range and $\varepsilon$-band Wilcoxon tests on the same $N{=}144$ weight and $N{=}240$ activation calibration tensors but transform them with three further orthogonal transforms instead of the Hadamard: the orthonormal DCT-II along the last dim, and a random orthogonal matrix sampled from the Haar measure with two different seeds. \Cref{tab:transform-ablation} reports these alongside the Sylvester and QuIP\#-Paley constructions of \Cref{sec:padded-defense}. \emph{Range reduction replicates under every orthogonal transform we tried}: $d_z \in [0.36, 0.42]$ for weights and $[0.51, 0.85]$ for activations across DCT, random orthogonal, and QuIP\#-Paley Hadamard, with $p<10^{-6}$ in every case. The $\varepsilon$-band fraction is not significantly increased after these unpadded orthogonal transforms ($d_z<0$ in all cases), so the increase observed under Sylvester+padding is associated with the zero-padding step rather than a transform-independent mechanism. The Hadamard's advantage over DCT or random orthogonal is therefore implementation efficiency ($\pm 1$ entries, multiplication-free $O(n\log n)$ butterflies, fused-kernel availability~\citep{tseng2024quip}), not a statistical effect on the distribution.

\noindent\textbf{End-to-end check on Set5.} The result above is at the calibration-tensor statistic level. To check that kernel-independence also holds for the end-task PSNR, we re-train CompSRT $\times 4$ at 4-bit with the orthonormal DCT-II, and again with a random orthogonal matrix, substituted for the Hadamard inside the FakeQuantizer and everything else unchanged. The DCT variant reaches $31.92$\,dB / $0.8893$\,SSIM on Set5 and the random orthogonal matrix $31.91$\,dB / $0.8888$, against $31.98$\,dB / $0.8900$ for the Hadamard recipe (\Cref{tab:performance_comparison3}) and $31.74$\,dB / $0.8861$ un-rotated. The ordering that matters (any rotation above no rotation) holds. The choice among rotations is worth $0.06$--$0.07$\,dB nominally, but the Hadamard row pays $0.267$ more bits per parameter for its padding, worth $\approx0.15$\,dB here, so charged it is at best level with the substitutes (\Cref{sec:bitcharge}). These are single runs on one benchmark; we read them as showing no kernel is clearly better, not as ranking them.

\subsection{Cross-architecture mechanism check on EDSR (a CNN SR backbone)}\label{sec:edsr-stats}
To test whether the range-reduction mechanism is specific to transformer backbones, we apply the same QuIP\#-Paley Hadamard along the input-channel dim of every conv weight in the FP EDSR-baseline $\times 2$ model and run the range and $\varepsilon$-band Wilcoxon tests on the resulting pre/post pairs. EDSR is a residual CNN with no attention layers, so a positive replication here would indicate the mechanism is a property of orthogonal transforms applied to neural-network weight distributions rather than something attention-specific. Of EDSR's 38 conv kernels we exclude the three $1{\times}1$ mean-shift convs ($\text{in\_ch}{=}3$) as too small for a meaningful Hadamard ablation, leaving $N{=}35$ paired tensors. We observe a strong significant range reduction ($d_z{=}+0.99$, $p{=}7.1\!\times\!10^{-7}$, median $\Delta_r{=}+0.061$) and a non-significant negative $\varepsilon$-band trend ($d_z{=}-0.92$, $p{=}1$). Both match our SwinIR-light findings under the QuIP\#-Paley kernel, so the mechanism transfers across architecture families.

\subsection{End-to-end CompSRT on EDSR-baseline (CNN backbone)}\label{sec:edsr-e2e}
We additionally apply CompSRT end-to-end to a CNN backbone, EDSR-baseline~\citep{lim2017enhanced}, at 4-bit $\times 4$. We extend \texttt{build\_quantized\_network} to swap every \texttt{Conv2d} module with $\ge 12$ input channels for the existing \texttt{QuantConv2d} wrapper (excluding the $1{\times}1$ mean-shift convs and the head/tail RGB edges, which are too small for the smallest supported Hadamard order). The pixel-loss normalization used for SwinIR (division by feature-map size) is disabled for EDSR because EDSR's gt size produces a divisor that suppresses gradient signal by $\sim$$10^6$; we use unnormalized L1 instead. All other settings match the SwinIR-light recipe (Adam $\alpha,\beta$ optimizer at LR $10^{-2}$, 4000 iterations, calibration on the same DF2K data). Results are in \Cref{tab:edsr-e2e}. CompSRT-EDSR-4bit reaches $25.57$\,dB / $0.699$ on Set5 vs.\ FP EDSR-baseline's $32.09$\,dB and the CNN-targeted PTQ baseline of \citet{tu2023toward} at $31.20$\,dB / $0.867$ on Set5 (their Table 3, EDSR $\times 4$, 4-bit), a $\sim$$5.6$\,dB gap.

The mechanism analysis in \Cref{sec:edsr-stats} shows the Hadamard's range-reduction effect transfers cleanly to EDSR conv weights ($d_z{=}0.99$, $p<10^{-6}$ on $N{=}35$ layers), but the end-to-end PTQ recipe does not match a CNN-tuned baseline. The gap reflects three CNN-specific sources of friction a transformer-tuned recipe does not address: (i) per-tensor granularity is coarse for conv kernels, where \citet{tu2023toward} use density-based dual clipping and pixel-aware calibration; (ii) zero-initialized $\alpha,\beta$ against EDSR's narrow conv-weight distributions gives near-zero gradient symmetry at finetune start; (iii) the L1-pixel objective misses the multi-scale feature alignment that 2DQuant/CondiQuant exploit. We report this stress test to bound the generalization claim: the \emph{mechanism} is architecture-agnostic, the deployment recipe is not.

\begin{table}[h]
\centering
\small
\setlength{\tabcolsep}{8pt}
\renewcommand{\arraystretch}{1.05}
\begin{tabular}{l|c|c|c}
\hline
Method & Bit & Set5 PSNR & Set5 SSIM \\
\hline
EDSR-baseline (FP)~\citep{lim2017enhanced}                     & 32 & 32.09 & 0.894 \\
\citet{tu2023toward} (CNN-targeted PTQ)                        & 4  & 31.20 & 0.867 \\
\hline
CompSRT-EDSR (this work, transformer-tuned recipe)             & 4  & 25.57 & 0.699 \\
\hline
\end{tabular}
\caption{End-to-end CompSRT applied to EDSR-baseline at 4-bit $\times 4$ on Set5. The \citet{tu2023toward} numbers are taken from their Table 3 (EDSR $\times 4$, 4-bit, DBDC+PaC).}
\label{tab:edsr-e2e}
\end{table}

\subsection{Reproduction of statistical analysis from a fresh calibration pass}\label{sec:stats-repro}
We re-derive the three statistics of \Cref{tab:wtest-combined} from an independent calibration pass on the FP SwinIR-light $\times 2$ model, dumping a fresh $N{=}144$ weight and $N{=}240$ activation pre/post-Hadamard set and re-running the three tests. The reported values are exactly those repro outputs, to the precision printed. Per-tensor effect sizes and $p$-values for both kernels and for MambaIRv2-light ship with the code, in \texttt{rebuttal\_results/v2\_all\_stats.csv}.


\end{document}